\documentstyle[twoside,epsf,amssymb]{article}

\catcode`\@=11
\long\def\@makefntext#1{
\protect\noindent \hbox to 3.2pt {\hskip-.9pt
$^{{\eightrm\@thefnmark}}$\hfil}#1\hfill}       

\def\@makefnmark{\hbox to 0pt{$^{\@thefnmark}$\hss}}    

\def\ps@myheadings{\let\@mkboth\@gobbletwo
\def\@oddhead{\hbox{}
\rightmark\hfil\eightrm\thepage}
\def\@oddfoot{}\def\@evenhead{\eightrm\thepage\hfil
\leftmark\hbox{}}\def\@evenfoot{}
\def\sectionmark##1{}\def\subsectionmark##1{}}

\oddsidemargin=\evensidemargin
\newcounter{sectionc}\newcounter{subsectionc}\newcounter{subsubsectionc}
\renewcommand{\section}[1] {\vspace{12pt}\addtocounter{sectionc}{1}
\setcounter{subsectionc}{0}\setcounter{subsubsectionc}{0}\noindent
    {\tenbf\thesectionc. #1}\par\vspace{5pt}}
\renewcommand{\subsection}[1] {\vspace{12pt}\addtocounter{subsectionc}{1}
\setcounter{subsubsectionc}{0}\noindent
{\bf\thesectionc.\thesubsectionc. {\kern1pt \bfit #1}}\par\vspace{5pt}}
\renewcommand{\subsubsection}[1] {\vspace{12pt}\addtocounter{subsubsectionc}{1}
    \noindent{\tenrm\thesectionc.\thesubsectionc.\thesubsubsectionc.
    {\kern1pt \tenit #1}}\par\vspace{5pt}}
\newcommand{\nonumsection}[1] {\vspace{12pt}\noindent{\tenbf #1}
    \par\vspace{5pt}}

\newcounter{appendixc}
\newcounter{subappendixc}[appendixc]
\newcounter{subsubappendixc}[subappendixc]
\renewcommand{\thesubappendixc}{\Alph{appendixc}.\arabic{subappendixc}}
\renewcommand{\thesubsubappendixc}
    {\Alph{appendixc}.\arabic{subappendixc}.\arabic{subsubappendixc}}

\renewcommand{\appendix}[1] {\vspace{12pt}
        \refstepcounter{appendixc}
        \setcounter{figure}{0}
        \setcounter{table}{0}
        \setcounter{lemma}{0}
        \setcounter{theorem}{0}
        \setcounter{corollary}{0}
        \setcounter{definition}{0}
        \setcounter{equation}{0}
        \renewcommand{\thefigure}{\Alph{appendixc}.\arabic{figure}}
        \renewcommand{\thetable}{\Alph{appendixc}.\arabic{table}}
        \renewcommand{\theappendixc}{\Alph{appendixc}}
        \renewcommand{\thelemma}{\Alph{appendixc}.\arabic{lemma}}
        \renewcommand{\thetheorem}{\Alph{appendixc}.\arabic{theorem}}
        \renewcommand{\thedefinition}{\Alph{appendixc}.\arabic{definition}}
        \renewcommand{\thecorollary}{\Alph{appendixc}.\arabic{corollary}}
        \renewcommand{\theequation}{\Alph{appendixc}.\arabic{equation}}
        \noindent{\tenbf Appendix \theappendixc #1}\par\vspace{5pt}}
\newcommand{\subappendix}[1] {\vspace{12pt}
        \refstepcounter{subappendixc}
        \noindent{\bf Appendix \thesubappendixc. {\kern1pt \bfit #1}}
    \par\vspace{5pt}}
\newcommand{\subsubappendix}[1] {\vspace{12pt}
        \refstepcounter{subsubappendixc}
        \noindent{\rm Appendix \thesubsubappendixc. {\kern1pt \tenit #1}}
    \par\vspace{5pt}}

\newcommand{\textlineskip}{\baselineskip=13pt}
\newcommand{\smalllineskip}{\baselineskip=10pt}

\newcommand{\copyrightheading}[1]
    {\vspace*{-2.5cm}\smalllineskip{\flushleft
    {\footnotesize Quantum Information and Computation, Vol.~1, No.~0 (2001) 000--000 #1}\\
    {\footnotesize \copyright\kern2pt Rinton Press}\\
     }}

\newcommand{\publisher}[2]{{\begin{center}\footnotesize\smalllineskip
    Received #1\\
    Revised #2
    \end{center}
    }}

\def\abstracts#1#2#3{{
    \centering{\begin{minipage}{4.5in}\footnotesize\baselineskip=10pt
    \parindent=0pt #1\par
    \parindent=15pt #2\par
    \parindent=15pt #3
    \end{minipage}}\par}}

\def\keywords#1{{
    \centering{\begin{minipage}{4.5in}\footnotesize\baselineskip=10pt
    {\footnotesize\it Keywords}\/: #1
     \end{minipage}}\par}}
\def\communicate#1{{
    \centering{\begin{minipage}{4.5in}\footnotesize\baselineskip=10pt
    {\footnotesize\it Communicated by}\/: #1
     \end{minipage}}\par}}

\renewenvironment{thebibliography}[1]
        {\frenchspacing
     \ninerm\baselineskip=11pt
         \begin{list}{\arabic{enumi}.}
        {\usecounter{enumi}\setlength{\parsep}{0pt}
     \setlength{\leftmargin 12.7pt}{\rightmargin 0pt}
         \setlength{\itemsep}{0pt} \settowidth
    {\labelwidth}{#1.}\sloppy}}{\end{list}}

\newcounter{itemlistc}
\newcounter{romanlistc}
\newcounter{alphlistc}
\newcounter{arabiclistc}

\newcommand{\fcaption}[1]{
        \refstepcounter{figure}
        \setbox\@tempboxa = \hbox{\footnotesize Fig.~\thefigure. #1}
        \ifdim \wd\@tempboxa > 5in
           {\begin{center}
        \parbox{5in}{\footnotesize\smalllineskip Fig.~\thefigure. #1}
            \end{center}}
        \else
             {\begin{center}
             {\footnotesize Fig.~\thefigure. #1}
              \end{center}}
        \fi}

\newcommand{\tcaption}[1]{
        \refstepcounter{table}
        \setbox\@tempboxa = \hbox{\footnotesize Table~\thetable. #1}
        \ifdim \wd\@tempboxa > 5in
           {\begin{center}
        \parbox{5in}{\footnotesize\smalllineskip Table~\thetable. #1}
            \end{center}}
        \else
             {\begin{center}
             {\footnotesize Table~\thetable. #1}
              \end{center}}
        \fi}

\def\pmb#1{\setbox0=\hbox{#1}
    \kern-.025em\copy0\kern-\wd0
    \kern.05em\copy0\kern-\wd0
    \kern-.025em\raise.0433em\box0}

\def\fnm#1{$^{\mbox{\scriptsize #1}}$}
\def\fnt#1#2{\footnotetext{\kern-.3em
    {$^{\mbox{\scriptsize #1}}$}{#2}}}

\def\fpage#1{\begingroup
\voffset=.3in
\thispagestyle{empty}\begin{table}[b]\centerline{\footnotesize #1}
    \end{table}\endgroup}

\def\runninghead#1#2{\pagestyle{myheadings}
\markboth{{\protect\footnotesize\it{\quad #1}}\hfill}
{\hfill{\protect\footnotesize\it{#2\quad}}}}
\font\tenrm=cmr10
\font\tenit=cmti10
\font\tenbf=cmbx10
\font\bfit=cmbxti10 at 10pt
\font\ninerm=cmr9

\font\eightrm=cmr8

\def\FigName{figure}%
\newbox\captionbox
\long\def\@makecaption#1#2{%
  \ifx\FigName\@captype
    \vskip\abovecaptionskip
    \setbox\tempbox\hbox{{\figurecaptionfont #1\hskip1em #2}}
    \ifdim\wd\tempbox< 28pc
    \centerline{\box\tempbox}
    \else
    {\figurecaptionfont #1\hskip1em #2\par}
\fi\else
    \setbox\tempbox\hbox{{\tablecaptionfont #1\hskip1em #2}}
    \ifdim\wd\tempbox< 28pc
    \centerline{\box\tempbox}
    \else
    {\tablecaptionfont #1\hskip1em #2\par}%
    \fi
 \vskip\belowcaptionskip
 \fi}
\InputIfFileExists{psfig.sty}
{\typeout{^^Jpsfig.sty inputed...ok}}{\typeout{^^JWarning:
psfig.sty could be be found.^^J}}
\InputIfFileExists{epsfsafe.tex}
{\typeout{^^Jepsfsafe.tex inputed...ok}}
            {\typeout{^^JWarning: epsfsafe.tex could not be found.^^J}}
\InputIfFileExists{epsfig.sty}
{\typeout{^^Jepsfig.sty inputed...ok}}{\typeout{^^JWarning: epsfig.sty could not be found.^^J}}
\InputIfFileExists{epsf.sty}
{\typeout{^^Jepsf.sty inputed...ok}}{\typeout{^^JWarning: epsf.sty could not be found.^^J}}%
\def\fps@figure{tbp}
\def\ftype@figure{1}
\def\ext@figure{lof}
\def\fnum@figure{Fig.\ \thefigure}
\def\qed{\hbox{${\vcenter{\vbox{              
   \hrule height 0.4pt\hbox{\vrule width 0.4pt height 6pt
   \kern5pt\vrule width 0.4pt}\hrule height 0.4pt}}}$}}

\begin{document}
\setlength{\textheight}{8.0truein}    

\runninghead{Optimization of Coherent Attacks   $\ldots$}
            {R.W. Spekkens and T. Rudolph}

\normalsize\textlineskip
\thispagestyle{empty}
\setcounter{page}{1}

\copyrightheading{} 

\vspace*{0.88truein}

\def\text#1{\mbox{\scriptsize #1}}

\fpage{1}
\centerline{\bf
Optimization of coherent attacks in generalizations of the BB84}
\vspace*{0.035truein}
\centerline{\bf quantum bit
commitment protocol}
\vspace*{0.37truein}
\centerline{\footnotesize
R. W. Spekkens\footnote{spekkens@physics.utoronto.ca }}
\vspace*{0.015truein}
\centerline{\footnotesize\it Department of Physics, University of Toronto, }
\baselineskip=10pt
\centerline{\footnotesize\it 60 St. George Street,
Toronto, Ontario, Canada, M5S 1A7}
\vspace*{10pt}
\centerline{\footnotesize T. Rudolph\footnote{rudolpht@bell-labs.com. Present address: 1D456 Bell Labs, Lucent Technologies, 700 Mountain Ave., Murray Hill, NJ 17974, U.S.A. }}
\vspace*{0.015truein}
\centerline{\footnotesize\it Institut f\"ur Experimentalphysik,
Universit\"at Wien, }
\baselineskip=10pt
\centerline{\footnotesize\it Boltzmanngasse 5, A-1090 Vienna, Austria}
\vspace*{0.225truein}
\publisher{(received date)}{(revised date)}

\vspace*{0.21truein}
\abstracts{
It is well known that no quantum bit
commitment protocol is unconditionally secure. Nonetheless, there
can be non-trivial upper bounds on both Bob's probability of
correctly estimating Alice's commitment and Alice's probability of
successfully unveiling whatever bit she desires. In this paper,
we seek to determine these bounds for generalizations of the BB84
bit commitment protocol.  In such protocols, an honest Alice
commits to a bit by randomly choosing a state from a specified
set and submitting this to Bob, and later unveils the bit to Bob
by announcing the chosen state, at which point Bob measures the
projector onto the state. Bob's optimal cheating strategy can be
easily deduced from well known results in the theory of quantum
state estimation. We show how to understand Alice's most general
cheating strategy, (which involves her submitting to Bob one half
of an entangled state) in terms of a theorem of
Hughston, Jozsa and Wootters. We also show how the problem of
optimizing Alice's cheating strategy for a fixed submitted state
can be mapped onto a problem of state estimation. Finally, using
the Bloch ball representation of qubit states, we identify the
optimal coherent attack for a class of protocols that can be implemented with just a single qubit. These results provide a
tight upper bound on Alice's probability of successfully
unveiling whatever bit she desires in the protocol proposed by
Aharonov {\it et al.}, and lead us to identify a qubit protocol
with even greater security. }{}{}

\vspace*{10pt}
\vspace*{3pt}
\communicate{to be filled by the Editorial}

\vspace*{1pt}\textlineskip  
\section{Introduction}
\label{I}           
\vspace*{-0.5pt}\noindent
Suppose Alice and Bob wish to play a game wherein Alice wins if
she can correctly predict which of two mutually exclusive events
will occur and Bob wins if she cannot. One way to play the game
would be for Alice to tell Bob her prediction before the events
in question. There are situations, however, where this is
inappropriate. For instance, Bob might be able to influence the
relative probability of the events in question (indeed, which of
these events occurs might be entirely up to Bob). In such cases,
Alice wants Bob to know as little as possible about her
prediction until some time after the occurrence of one of the
events. Of course, Bob will still want to receive
some sort of `token' of Alice's prediction prior to the events in
question, since otherwise Alice could always claim to have won
the game. Thus, Alice and Bob would like a cryptographic protocol
which forces Alice to `commit' herself to a bit (which encodes
her prediction), while ensuring that Bob can find out as little
as possible about this bit until the time that Alice reveals it to
him. This is a bit commitment(BC) protocol. In addition to the
task of prediction described above, BC appears as a primitive in
many other cryptographic tasks and is therefore of particular
significance in cryptography.

A simple example of an implementation of BC proceeds as follows.
Alice writes a `0' or a `1' on a piece of paper, and locks this
in a safe. She then sends the safe to Bob, but keeps the key.
When it comes time to reveal her commitment, she sends the key to
Bob, who opens the safe and discovers the value of the bit. This
protocol binds Alice to the bit she chose at the outset since she
cannot change what is written on the piece of paper after she
submits the safe to Bob. However, it only conceals the bit from
Bob if he is unable to pick the lock, or force the safe open, or
image the contents of the safe.

This paper focuses on a particular class of quantum BC protocols,
specifically, generalizations of the BC protocol that was
published by Bennett and Brassard in 1984~\cite{BB84}. We shall
refer to these as {\em generalized BB84 BC protocols}. Within
such protocols, an honest Alice commits to a bit 0 by choosing a
state randomly from a specified set of states, and by
subsequently sending a system prepared in this state to Bob. She
commits to a bit 1 by choosing the state from a different set. At
the end of the protocol she reveals to Bob which state she
submitted and Bob measures the projector onto this state to
verify Alice's claim.

Bob can cheat in such a protocol by performing a measurement on
the systems submitted to him by Alice, prior to Alice revealing
her commitment. The measurement that maximizes his probability of
correctly estimating Alice's
commitment can be determined from the well-known theory of state estimation~\cite{Helstrom}.

Alice can cheat by preparing the system she initially submits to
Bob in a state different from the ones specified by the protocol,
in particular, by entangling this system with an ancilla system
that she keeps in her possession, and by later performing a
measurement on the ancilla and choosing the state which she
announces to Bob based on the outcome of this measurement. This
has been called a {\em coherent attack}, since in general such an
attack requires Alice to maintain the coherence between the
different possibilities in the random choice the protocol asks
her to make. It has also been called an {\em EPR-type attack},
since in the original BB84 BC protocol, the optimal entangled
state for Alice to prepare is the EPR state. The problem of
determining the coherent attack that maximizes Alice's
probability of successfully cheating has remained open to date.
It is the goal of this paper to begin to answer this question.

It has been shown by Mayers~\cite{Mayers} and by Lo and
Chau~\cite{LC} that an unconditionally secure BC protocol does not
exist~\cite{Localitycaveat}. In
other words, it is not possible to devise a BC protocol that is {\em %
arbitrarily concealing}, that is, one for which Bob's
probability of
correctly estimating Alice's commitment is arbitrarily small, and {\em %
arbitrarily binding}, that is, one for which Alice's
probability of revealing whatever bit she desires without being
caught cheating is arbitrarily small. Nonetheless, there remain
interesting questions to be answered about coherent attacks. For
instance, it {\em is }possible to have
a BC\ protocol that is {\em partially binding} and {\em partially concealing}%
, wherein Alice and Bob's probabilities of successfully cheating
are both bounded above~\cite{SpekkensRudolph2}. Determining the
optimal coherent attack is crucial to determining the degree of
bindingness that can be achieved in such protocols.

Coherent attacks are also important in other quantum
cryptographic tasks between mistrustful parties - such as coin
tossing~\cite{LCcointossing}, cheat sensitive bit
commitment~\cite{KentandHardy}, bit escrow~\cite{Aharanov} and
quantum gambling~\cite{Goldenberg} - wherein a type of bit
commitment often appears as a subprotocol. Understanding how to
optimize coherent attacks is therefore important for settling
questions about the degree of security that can be achieved for
such tasks.

We summarize here the main results of the paper. The last four
apply only to protocols that can be implemented using a single
qubit.

\begin{itemize}
\item  We explain coherent attacks in terms of the well known theorem of
Hughston, Jozsa and Wootters~\cite{Hughston Josza and Wooters}.

\item  We demonstrate that the problem of finding the optimal coherent
attack for a {\em fixed }submitted state can be mapped onto a
problem of state estimation which has a known solution \cite{Helstrom}.

\item  We show that the optimal state for a cheating Alice to submit has a support in the span of the supports of the set of states from which an honest Alice chooses.

\item  We provide a simple geometrical picture on the Bloch sphere of
coherent attacks. In addition to being useful for building one's
intuitions about such attacks, this provides a convenient
formalism within which to solve the optimization problem, as well
as a geometrical criterion for whether or not Alice can cheat
with probability 1 in a given protocol.

\item  We find analytic expressions for the optimal cheating strategy in the
case where the sets of states that an honest Alice chooses from
each have no more than two elements.

\item  Using these results, we determine Alice's optimal coherent attack in
a BC protocol that was proposed by Aharonov {\em et
al.}~\cite{Aharanov}. Our result provides a tight upper bound on
Alice's probability of unveiling whatever bit she desires,
improving upon the best previous known upper bound. This allows
us to determine, for this protocol, the trade-off relation
between a measure of the concealment and a measure of the bindingness.
We show that the same trade-off relation can be achieved with
several other protocols.

\item  Finally, our results allow us to determine Alice's optimal coherent
attack in a novel generalized BB84 BC protocol wherein the
trade-off relation between concealment and bindingness is better
than can be achieved with the protocol of Aharonov {\em et al.}
\end{itemize}

  The paper is organized as follows. In section~2, we provide an
operational definition of bit commitment, define degrees of
security, and describe the BB84 BC protocol and its
generalizations. In section~3, we introduce the notion of a
convex decomposition of a density operator, review its
properties, and demonstrate its significance for coherent
attacks. In section~4, we formulate the optimization problem to
be solved. Results for protocols involving systems of arbitrary
dimensionality and for protocols involving qubits are presented
in sections~5 and 6 respectively. Applications of these results
are presented in section~7, and section~8 contains our
concluding remarks.

\section{Bit Commitment}
\label{II}

\subsection{An operational definition of Bit Commitment}
\label{IIA}
\noindent
We begin by providing a definition of BC that is strictly
operational, that is, one which only makes reference to the
experimental operations carried out by the parties, and not to
any concepts that are particular to a physical theory. This seems
to us to be the most sensible way of proceeding for {\em any}
information processing task, since such tasks can be defined
independently of their physical implementation and consequently
of any physical theory describing this implementation. Among
other benefits, this approach allows one to characterize a
physical theory by the type of protocols which can be securely
implemented within a universe described by that theory.

A BC protocol is a cryptographic protocol between two mistrustful
parties.
It can be defined in terms of the characteristics of these parties' honest (%
i.e., non-cheating) strategies. We call the two parties
Alice and Bob, and assume that Alice is the one making the
commitment.

  The protocol is divided into three intervals, called the
commitment phase, the holding phase and the unveiling phase. Each
of these may involve many rounds of communication between Alice
and Bob. The result of the protocol is one of three
possibilities, denoted `0', `1' and `fail'. Which of these has
occurred is determined from the outcomes of all the measurements
that an honest Bob has made throughout the protocol. The protocol
specifies the strategy an honest Alice must adopt to commit to a
bit $b.$ It is such that if both parties are honest and Alice
follows the strategy for committing a bit $0(1),$ the result of
the protocol is necessarily `0'(`1'). It follows that if the
outcome `fail' occurs, an honest Bob can conclude that Alice must
have cheated. The protocol must also be such that if both parties
are honest, Alice does not, through actions taken after the end
of the commitment phase, change the relative probability of the
results `0' and `1' occurring, and Bob does not, prior to the
beginning of the unveiling phase, gain any information about
Alice's commitment.

In the protocols we shall be considering, Alice will not always
be caught
when she cheats. Thus, it can happen that the result of the protocol is $`b$%
' even though Alice cheated and did not follow the honest
strategy for committing a bit $b.$ Indeed, Alice can, by
cheating, change the relative probability of the `0' and `1'
results by actions taken after the commitment phase. We shall say that `Alice unveils bit $b$'
whenever the result of the protocol is $`b.$'\fnm{a}\fnt{a}{ It is important
to remember that within our terminology `Alice unveiling bit $b$'
implies that she was not caught cheating. Thus in a generalized
BB84 BC
protocol, when Alice announces $b$ to Bob$,$ we say that Alice is {\em %
attempting} to unveil a bit $b,$ but we only say that she has
unveiled $b$ if she passes Bob's test.}

\subsection{\protect  Types of security}
\label{IIB}
\noindent
To define the security of a BC protocol, one needs to quantify
the notions of concealment against Bob and bindingness against
Alice. In this paper, we focus upon the probability that Bob can,
prior to the beginning of the unveiling phase, correctly estimate
Alice's commitment (given that Alice is honest), and the
probability that Alice can, after the end of the commitment
phase, successfully unveil whatever bit she desires (given that
Bob is honest). We denote these by $P_{E}$ and $P_{U}$
respectively. Note that these probabilities vary with the
cheating strategy used. In this paper, we shall only consider
protocols wherein these are both equal to 1/2 for honest
strategies.\fnm{b}\fnt{b}{In most discussions of bit commitment,
it is assumed that neither Alice nor Bob has any information at
the commitment phase about which bit will be more beneficial for
Alice to unveil. However, one must relax this assumption in order
to consider a game wherein Alice predicts which of two events
will occur given some prior information on their relative
probability. The results of this paper can be generalized in a
straightforward manner to
apply to such a protocol. It suffices to replace Eq.~(\ref{P_U}) with $%
P_{U}=p_{0}P_{U0}+p_{1}P_{U1},$ where $p_{b}$ is the probability
that Alice will wish to unveil bit $b$ after the commitment
phase, and to generalize all subsequent expressions accordingly.}

A bit commitment protocol is said to be {\em arbitrarily binding}
if for all of Alice's strategies, $P_{U}$ is bounded above by
$1/2+\varepsilon ,$ where $\varepsilon $ can be made arbitrarily
small by increasing some security parameter in the protocol. It
is said to be {\em arbitrarily concealing }if for all of Bob's
strategies, $P_{E}$ is bounded similarly. Although no BC protocol can be arbitrarily binding and
arbitrarily concealing, both $P_{E}$ and $P_{U}$ {\em can }have
non-trivial upper bounds (that is, upper bounds less than $1)$.
We will refer to such protocols as {\em partially binding }and
{\em partially concealing}. The
maxima of $P_{E}$ and $P_{U}$ for a given protocol, which we denote by $%
P_{E}^{\max }$ and $P_{U}^{\max }$, quantify the degree of
concealment and the degree of bindingness that can be achieved in
this protocol.

The implementation of BC using a safe, discussed in the
introduction, is binding against Alice, but is only concealing
against Bob if he has limited `safe-cracking' resources. More
useful implementations of bit commitment instead rely for
concealment on the assumption that Bob has limited computational
resources. Obviously, one would prefer that the security of the
protocol not depend on the resources of either party, but rather
only on the laws of physics and the integrity of the party's
laboratories. A
property of a protocol that has this feature is said to hold {\em %
unconditionally. }All the properties of protocols referred to in
this paper, are properties which hold unconditionally.

\subsection{The BB84 BC protocol}
\label{IIC}
\noindent
The first proposal for a quantum mechanical implementation of a
BC\ protocol was made by Bennett and Brassard~\cite{BB84}. We
refer to it as the BB84 BC protocol. This was recognized by its
authors to have no bindingness against Alice. Nonetheless, we
begin by reviewing this protocol, since it provides a simple
example of the type of cheating strategy with which this paper
will be concerned.

Imagine a protocol wherein Alice submits a qubit to Bob during the
commitment phase. To commit to a bit 0, she prepares the qubit in
a state chosen uniformly from the set $\left\{ \left|
0\right\rangle ,\left| 1\right\rangle \right\} $, while to commit
to a bit 1, she chooses from the
set $\left\{ \left| +\right\rangle ,\left| -\right\rangle \right\} ,$ where $%
\left| \pm \right\rangle \equiv \left( \left| 0\right\rangle \pm \left|
1\right\rangle \right) /\sqrt{2}.$ No measurement Bob can do is
able to distinguish these two possibilities. At the unveiling
phase, Alice can tell Bob which state she submitted, and Bob can
do a measurement of the projector onto this state to verify her
honesty. If Alice tries to convince Bob that she submitted a
state drawn from the opposite set - for instance, that she
submitted $\left| +\right\rangle $ when in fact she submitted
$\left| 0\right\rangle $ - then her probability of passing his
test is only $1/2.$ The BB84 BC protocol demands that Alice
repeat her commitment for $N$ qubits, that is, that Alice either
chooses each qubit's state uniformly from $\left\{ \left|
0\right\rangle ,\left| 1\right\rangle \right\} $ or uniformly
from $\left\{ \left| +\right\rangle ,\left| -\right\rangle
\right\} $. Clearly, in this case her probability of passing Bob's test when
she lies about her commitment is $1/2^{N}$. So, with respect to
strategies wherein Alice cheats by lying about her commitment,
such a protocol appears to be arbitrarily binding.

However, Alice has another cheating strategy available to her.
Prior to submitting a qubit to Bob, she can entangle it with a
qubit that she keeps
in her possession. Specifically, she prepares the two in the EPR state $%
\left( \left| 0\right\rangle \left| 1\right\rangle -\left| 1\right\rangle
\left| 0\right\rangle \right) /\sqrt{2}.$ Given that this state can also be
written as $\left( \left| +\right\rangle \left| -\right\rangle
-\left| -\right\rangle \left| +\right\rangle \right) /\sqrt{2},$
it is clear that by measuring the $\left\{ \left| 0\right\rangle
,\left| 1\right\rangle \right\} $ basis or $\left\{ \left|
+\right\rangle ,\left| -\right\rangle \right\} $ basis on the
qubit in her possession, she projects the qubit in Bob's
possession into the $\left\{ \left| 0\right\rangle ,\left|
1\right\rangle
\right\} $ basis or $\left\{ \left| +\right\rangle ,\left| -\right\rangle
\right\} $ basis respectively. Moreover, the binary outcome of her
measurement will be perfectly anti-correlated with the state of
Bob's qubit. So Alice knows precisely which state to announce to
Bob. Using this strategy, she can choose which bit she wants to
unveil just prior to the unveiling phase, and always succeed at
passing Bob's test. This is the so-called `coherent' or `EPR'
attack.

The analysis thus far leaves open the possibility that some other
protocol using quantum primitives might succeed where the BB84
protocol failed. In fact, it has been shown that for the most
general nonrelativistic protocol, unconditional security is not
possible~\cite{Mayers,LC,LCcointossing}. Nonetheless,
there exist simple generalizations of the BB84 protocol that are
both partially concealing and partially binding.

\subsection{Generalizations of the BB84 BC protocol}
\label{IID}
\noindent
A generalized BB84 BC protocol defines two sets of states
$\left\{ \psi _{k}^{0}\right\} _{k=1}^{n_{0}}$ and $\left\{ \psi
_{k}^{1}\right\} _{k=1}^{n_{1}}$ and corresponding probability
distributions $\left\{ p_{k}^{0}\right\} _{k=1}^{n_{0}}$ and
$\left\{ p_{k}^{1}\right\} _{k=1}^{n_{1}}$ (note that the values
of $n_{0}$ and $n_{1}$ need not be the
same). In order to commit to bit $b,$ an honest Alice chooses a state from $%
\{\psi _{k}^{b}\}_{k=1}^{n_{b}}$ using the distribution $\left\{
p_{k}^{b}\right\} _{k=1}^{n_{b}}$ and sends a system prepared in
this state to Bob at the commitment phase. An honest Bob simply
stores the system
during the holding phase. At the unveiling phase, an honest Alice announces $%
b$ and $k$ to Bob, and he measures the projector onto $\left| \psi
_{k}^{b}\right\rangle .$ If Alice passes Bob's test, she has
succeeded in unveiling the bit $b,$ and the result of the
protocol is `$b$'$.$ Otherwise, she is caught cheating, and the
result of the protocol is $`$fail'.\fnm{c}\fnt{c}{ It should be
noted that the honest strategy for Alice to commit $b$ that we
have described is equivalent with respect to concealment to the
following strategy: Alice couples the system she sends to Bob
with a system she keeps in her possession (of dimension $n_{b}$
or greater) such that the two are in the entangled state
$\sum_{k=1}^{n_{b}}(p_{k}^{b})^{1/2}\left| k\right\rangle
\otimes \left| \psi _{k}^{b}\right\rangle ,$ where the $\left|
k\right\rangle $ form an orthonormal basis. At the unveiling
phase, she measures the basis $\left| k\right\rangle $ in order
to determine what integer to announce to Bob.}

  To estimate Alice's commitment, Bob must estimate whether the
system in his possession is described by $\rho
_{0}=\sum_{k=1}^{n_{0}}p_{k}^{0}\left| \psi _{k}^{0}\right\rangle
\left\langle \psi _{k}^{0}\right| ,$ or $\rho
_{1}=\sum_{k=1}^{n_{1}}p_{k}^{1}\left| \psi _{k}^{1}\right\rangle
\left\langle \psi _{k}^{1}\right| .$ The problem of optimal state estimation
has previously been studied in great
detail~\cite{Helstrom}, and in particular the optimal
measurement for discriminating two density operators is well
known~\cite{Fuchs}. Using the optimal measurement, the maximum probability of
Bob correctly estimating Alice's commitment is
\begin{equation}
P_{E}^{\max }=\frac{1}{2}+\frac{1}{4}{\rm Tr}\left| \rho _{0}-\rho
_{1}\right| ,  \label{Helstrom}
\end{equation}
where $\left| A\right| =\sqrt{A^{\dag }A}.$ It follows that as
long as $\rho _{0}$ and $\rho _{1}$ do not have orthogonal
supports, $P_{E}^{\max }$ is strictly less than 1 and the
protocol is partially concealing.

The complementary problem, of determining $P_{U}^{\max}$, the maximum probability of Alice unveiling whatever bit she desires, and the strategy
which achieves this maximum, has remained open to date. Alice's
most general strategy is of the following form. Prior to sending
the system to Bob, she entangles it with a system she keeps in
her possession. At the unveiling phase, she does one of two
measurements on the system in her possession, depending on
whether she is attempting to unveil a $0$ or a $1.$ She chooses
what integer $k$ to announce to Bob based on the outcome of this
measurement. It follows that in order to determine $P_{U}^{\max
}$, we must optimize over the entangled state that Alice
prepares, the two measurements she can perform and the
announcement she makes to Bob given each possible outcome.
We shall see that there exist generalized BB84 BC protocols where $%
P_{U}^{\max }$ is strictly less than 1, so that these protocols
are partially binding.

It will be useful to introduce a few mathematical concepts and
results before turning to the optimization problem.

\section{ Convex decompositions of a density operator}
\label{III}
\subsection{\protect  Definition and properties of convex
decompositions}
\label{IIIA}
\noindent
We begin by introducing a mathematical concept that will be
critical for solving our problem. A {\em convex decomposition}
$\left\{ \left( q_{k},\sigma _{k}\right) \right\} _{k=1}^{n}$ of
a density operator $\rho $ is a set of probabilities, $q_{k},$
and distinct density operators, $\sigma _{k},$ such that
\[
\rho =\sum_{k=1}^{n}q_{k}\sigma _{k}.
\]
The $\sigma _{k}$ will be referred to as the {\em elements }of
the convex decomposition. We use the term `convex' to distinguish
this from a decomposition of a pure state into a sum of pure
states, and from a decomposition of a density operator into
general sums of operators, that is, sums of operators that are
not necessarily positive. Nonetheless, we will throughout this
paper use the term {\em decomposition }as a
shorthand.\fnm{d}\fnt{d}{ Note that previous authors have used
the term $\rho $-{\em ensemble} to refer to a convex
decomposition of $\rho.$}

Some terminology will be used in connection with convex
decompositions. The
elements that receive non-zero probability will be called the {\em %
positively-weighted} elements. A decomposition will be called {\em extremal }%
if its positively-weighted elements are all of rank 1 (i.e., if
they are all pure states). A set of density operators will be
called {\em uncontractable} if none of its members can be written
as a convex decomposition of the others. A convex decomposition
will be called uncontractable if its positively-weighted elements
are uncontractable. Clearly, all extremal{\em \ }decompositions
are uncontractable. Finally, a{\em \ }decomposition of $\rho $ is
{\em trivial }if its only positively-weighted element is $\rho .$

Another concept that will be useful in the present investigation
is a relation that holds between sets of density operators, and
which we shall
refer to as {\em composable coincidence. }Two sets of density operators $%
\{\sigma _{k}^{0}\}$ and $\{\sigma _{k}^{1}\}$ will be called composably
coincident if there exist probability distributions $\left\{
q_{k}^{0}\right\} $ and $\left\{ q_{k}^{1}\right\} $ such that
\[
\sum_{k}q_{k}^{0}\sigma _{k}^{0}=\sum_{k}q_{k}^{1}\sigma _{k}^{1}.
\]
In other words, $\{\sigma _{k}^{0}\}$ and $\{\sigma _{k}^{1}\}$
are composably coincident if there exists a density operator
which has a convex decomposition in terms of the $\sigma
_{k}^{0}$'s and a convex decomposition in terms of the $\sigma
_{k}^{1}$'s$.$

It will also be useful to set forth a few well-known facts about
convex decompositions~\cite{Hughston Josza and Wooters}. A
necessary and sufficient condition for a density operator $\sigma
$ to appear in some convex decomposition of $\rho $ is for the
eigenvectors of $\sigma $ to be confined
to the support of $\rho .$ The cardinality of an extremal decomposition of $%
\rho $ must be greater than or equal to the rank of $\rho $. Finally, there
sometimes exists a prescription for obtaining the probability
with which a particular element appears in a convex decomposition
of a density operator. In convex decompositions of $\rho $
containing {\em orthogonal }elements, the probability associated
with an element $\sigma $ is fixed by $\rho $ and $\sigma $ -- it
is simply ${\rm Tr}(\sigma \rho )/{\rm Tr}\left( \sigma
^{2}\right).$ However, for a general set of non-orthogonal elements $%
\{\sigma _{k}\}$ that form a convex decomposition of $\rho ,$ the
probabilities need not be unique; the same set of density operators $%
\{\sigma _{k}\}$ may appear in different convex decompositions of $\rho .$
For instance, the completely mixed state in a 2d Hilbert space,
$I/2,$ has
an indenumerably infinite number of convex decompositions with elements $%
\left\{ \left| 0\right\rangle \left\langle 0\right| ,\left| 1\right\rangle
\left\langle 1\right| ,\left| +\right\rangle \left\langle +\right| ,\left|
-\right\rangle \left\langle -\right| \right\} $, since these
yield a
decomposition for every probability distribution of the form $\left( \frac{1%
}{2}\lambda ,\frac{1}{2}\lambda ,\frac{1}{2}(1-\lambda ),\frac{1}{2}%
(1-\lambda )\right) $ where $\lambda $ lies between $0$ and $1.$
Nonetheless, a special case wherein the probabilities {\em are
}unique is if the convex decomposition is extremal and of
cardinality equal to the rank of $\rho $. In this case, a simple
formula for the probability of a given element can be given. If
$\left\{ \left( q_{k},\left| \xi _{k}\right\rangle \left\langle
\xi _{k}\right| \right) \right\} $ is such a decomposition, then the
non-zero probabilities are given by Jaynes' rule~\cite{Jaynes},

\begin{equation}
q_{k}=\frac{1}{\left\langle \xi _{k}\right| \rho ^{-1}\left| \xi
_{k}\right\rangle },  \label{Jaynesrule}
\end{equation}
where $\rho ^{-1}$ is the inverse of the restriction of $\rho $
to its support (in other words, $\rho ^{-1}$ is obtained from $\rho $ by inverting the non-zero eigenvalues in the spectral resolution of $\rho $).

\subsection{The connection between convex decompositions and POVMs}
\label{IIIB}
\noindent
  The most general measurement on a system in quantum mechanics is
associated with a {\em positive operator-valued measure(POVM)}. A
POVM is a
set of positive operators that sum to the identity operator, that is, a set $%
\{E_{k}\}$ such that for every $k,$ $\left\langle \phi \right| E_{k}\left|
\phi \right\rangle \ge 0$ for all $\left| \phi \right\rangle \in {\cal H},$
and $\sum_{k}E_{k}=I.$ Neumark's theorem~\cite{Peres Neumark}
shows that every POVM on a system can be implemented by coupling
to an ancilla system and performing projective measurements on
the ancilla. As it turns out,
there is a close mathematical connection between convex decompositions of $%
\rho $ and POVMs, as was demonstrated by Hughston, Jozsa and Wootters~\cite
{Hughston Josza and Wooters}.

\begin{description}
\item[Lemma]  \quad There is a one-to-one map between the convex
decompositions of $\rho $ and the POVMs over the support of $\rho
.$ Specifically, the POVM $\{E_{k}\}_{k=1}^{n}$ is associated
with the decomposition $\{\left( q_{k},\sigma _{k}\right)
\}_{k=1}^{n}$ defined by
\begin{equation}
q_{k}\sigma _{k}=\sqrt{\rho }E_{k}\sqrt{\rho }.
\label{connection}
\end{equation}
\end{description}

{\bf Proof}. It is trivial to see that $\{(q_{k},\sigma
_{k})\}_{k=1}^{n}$ is a decomposition of $\rho $ by summing Eq.~\ref{connection} over $k$ and using the fact that
$\sum_{k}E_{k}=I$, where $I$ is the identity operator on the support of $\rho .$ That {\em any }decomposition of $\rho $ is
associated with {\em some} POVM over the support of $\rho$ follows from the fact that
$\sqrt{\rho } $ is invertible on the support of $\rho .$
Specifically, if this inverse is denoted by $\rho ^{-1/2}$ then
the resolution $\{(q_{k},\sigma _{k})\}_{k=1}^{n}$ is associated
with the POVM $\{E_{k}\}_{k=1}^{n}$ defined by $E_{k}=q_{k}\rho
^{-1/2}\sigma _{k}\rho ^{-1/2}.$ $\Box$

We say that the POVM $\{E_{k}\}_{k=1}^n$ {\em generates
}the convex decomposition $\{(q_{k},\sigma _{k})\}_{k=1}^{n}$.
Note that we do not treat the technicalities associated with
decompositions of infinite cardinality in this paper, however a
discussion of these can be found in Cassinelli {\em et
al.}~\cite{Cassinelli}.

\subsection{The significance of convex decompositions to coherent attacks}
\label{IIIC}
\noindent
Suppose Alice and Bob share an entangled state for which $\rho $
is the reduced operator on Bob's system. Prior to any
measurements, the best Alice can do in predicting the outcomes of
Bob's measurements is to use the density operator $\rho $ in the
Born rule. However, by virtue of the correlations between her
system and Bob's, if she performs a measurement and takes note of
the outcome, her ability to predict the outcomes of Bob's
measurements will increase. Since all of the information that is
relevant to Alice predicting the outcomes of Bob's measurements
is encoded in a density operator, it follows that when she learns
the outcome of her measurement, she should update the density
operator with which she describes Bob's
system. Suppose that the $k$th outcome occurs with relative frequency $q_{k}$%
, and leads Alice to update the density operator with which she
describes Bob's system to $\sigma _{k}.$ We say that the
statistics of possible updates of Alice's description of Bob's
system are given by $\left\{ \left( q_{k},\sigma _{k}\right)
\right\} ,$ that is, a set of probabilities and density
operators$.$

As it turns out, the possibilities for these statistics are given
by the convex decompositions of $\rho .$ Specifically, we have:

\begin{description}
\item[HJW Theorem]  \quad For {\em every} measurement Alice can perform, the
statistics of possible updates of her description is given by
{\em some} convex decomposition of $\rho ,$ and for {\em every
}convex decomposition of $\rho ,$ there exists {\em some}
measurement for which the statistics of possible updates is given
by that decomposition.
\end{description}

This was first demonstrated for extremal convex decompositions by
Hughston, Jozsa and Wootters~\cite{Hughston Josza and Wooters},
and it is straightforward to generalize the proof to arbitrary
convex decompositions. Since this theorem is the key to coherent
attacks, we present the generalized proof here.

  {\bf Proof. }Suppose Alice and Bob share a state $\left| \psi
\right\rangle $ that is a purification of $\rho $ (a normalized vector in ${\cal H}%
_{A}\otimes {\cal H}_{B}$ satisfying ${\rm Tr}_{A}\left( \left| \psi
\right\rangle
\left\langle \psi \right| \right) =\rho $). If the non-zero eigenvalues of $%
\rho $ are denoted by $\lambda _{j},$ and $\{\left| e_{j}\right\rangle \}$
is a set of normalized eigenvectors associated with these eigenvalues, then $%
\left| \psi \right\rangle $ can always be written as
\[
\left| \psi \right\rangle =\sum_{j}\sqrt{\lambda _{j}}\left|
f_{j}\right\rangle \otimes \left| e_{j}\right\rangle ,
\]
where $\{\left| f_{j}\right\rangle \}$ is a set of orthonormal
vectors for Alice's system. This way of writing $\left| \psi
\right\rangle $ is known as the bi-orthogonal or Schmidt
decomposition.

We begin by specifying the measurement that Alice must do on her
system in order to have her statistics of possible updates given
by the convex decomposition $\left\{ \left( q_{k},\sigma
_{k}\right) \right\} $ of $\rho .$
If the POVM on Bob's system that generates this decomposition is denoted by $%
\{E_{k}\},$ so that Eq.~(\ref{connection}) holds, and $U$ is the unitary map that satisfies
\[
\left| f_{j}\right\rangle =U\left| e_{j}\right\rangle ,
\]
then the required measurement on Alice's system is the one
associated with the POVM $\{UE_{k}^{T}U^{\dag }\}_{k=1}^{n}$, where $E_{k}^{T}$ denotes the transpose of $E_{k}$ with respect to the basis of eigenvectors of $\rho$ (note that this POVM need only be defined over the support of ${\rm Tr}_{B}\left( \left| \psi
\right\rangle \left\langle \psi \right| \right)$). The
proof is as follows.

The entangled state Alice and Bob share can be written in terms
of $U$ as
\[
\left| \psi \right\rangle =\sum_{j}\sqrt{\lambda _{j}}U\left|
e_{j}\right\rangle \otimes \left| e_{j}\right\rangle .
\]
Upon measuring the POVM $\{UE_{k}^{T}U^{\dag }\}$ on her system and
obtaining outcome $k,$ the projection postulate for POVMs
dictates that Alice should describe Bob's system by the
unnormalized state
\begin{eqnarray*}
{\rm Tr}_{A}\left( \sqrt{UE_{k}^{T}U^{\dag }}\left| \psi
\right\rangle
\left\langle \psi \right| \sqrt{UE_{k}^{T}U^{\dag }}\right)
&=&\Big( \sum_{j}\sqrt{\lambda _{j}}\left| e_{j}\right\rangle
\left\langle
e_{j}\right| \Big) E_{k}\Big( \sum_{j^{\prime }}\sqrt{\lambda _{j^{\prime }}}%
\left| e_{j^{\prime }}\right\rangle \left\langle e_{j^{\prime }}\right| \Big)
\\
&=&\sqrt{\rho }E_{k}\sqrt{\rho } \\
&=&q_{k}\sigma _{k}.
\end{eqnarray*}
So after this measurement, with probability $q_{k}$ Alice updates the
density operator with which she describes Bob's system to $\sigma
_{k}$.

It is also easy to show that the statistics of possible updates
are given by {\em some} convex decomposition of $\rho $ for {\em
every }measurement Alice can do. This follows from the fact that every
POVM can be written in the form $\{UE_{k}^{T}U^{\dag }\}$ for some choice of $\{E_{k}\}$
given a particular $U.$ $\Box$

When Alice entangles the system she submits to Bob with a system
she keeps in her possession in such a way that Bob's reduced
density operator is $\rho ,$ we shall say that Alice {\em submits
}$\rho $ to Bob. When Alice performs a measurement that leads to
her statistics of possible updates being given by the convex
decomposition $\left\{ \left( q_{k},\sigma _{k}\right)
\right\} $ of $\rho ,$ we shall say that Alice {\em realizes }this
decomposition on Bob's system.

\section{The nature of the optimization problem}
\label{IV}
\noindent
In section~2.4, we formulated the problem of determining the
optimal cheat strategy for Alice as a variational problem over
the entangled state that she initially prepares and the
measurements she performs on her half of the system. However,
from the results of section~3.3 it is clear that in determining
Alice's probability of unveiling the bit of her choosing, all
that is important about the entangled state she prepares is the
reduced density operator $\rho $ she submits to Bob, and all that
is important about the measurement she performs is the convex
decomposition of $\rho $ that she thereby realizes. It suffices
therefore to vary over $\rho $ and its convex decompositions.

We begin by showing that if Alice is attempting to unveil a bit
$b$ then it suffices for her to realize a convex decomposition
with a number of elements less than or equal to $n_{b}.$ The
proof is as follows. Suppose Alice
realizes a convex decomposition $\left\{ \left( \tilde{q}_{j},\tilde{\sigma}%
_{j}\right) \right\} _{j=1}^{n^{\prime }}$ with a number of elements $%
n^{\prime }$ that is greater than $n_{b}$. She still must
announce to Bob an index between $1$ and $n_{b},$ so that the
elements of this decomposition must be grouped into $n_{b}$ sets,
where elements in the $k$th set, $S_{k},$
correspond to announcing the index $k$ to Bob. When Alice announces index $k$%
, Bob will measure the projector $\left| \psi
_{k}^{b}\right\rangle
\left\langle \psi _{k}^{b}\right| $ and obtain a positive result with
probability $\sum_{j\in S_{k}}\tilde{q}_{j}\left\langle \psi
_{k}^{b}\right|
\tilde{\sigma}_{j}\left| \psi _{k}^{b}\right\rangle .$ However, there is
always an $n_{b}$-element convex decomposition that yields the
same probability of a positive result as the one considered here;
specifically, the decomposition $\left\{ \left( q_{k},\sigma
_{k}\right) \right\}
_{k=1}^{n_{b}}$ with $q_{k}\sigma _{k}=\sum_{j\in S_{k}}\tilde{q}_{j}\tilde{%
\sigma}_{j}.$

The probability of Alice succeeding at unveiling the bit $b$
given that she submits $\rho $ and realizes an $n_{b}$-element convex
decomposition $\left\{
\left( q_{k},\sigma _{k}\right) \right\} _{k=1}^{n_{b}}$ of $\rho $ is
\begin{equation}
P_{Ub}=\sum_{k=1}^{n_{b}}q_{k}\left\langle \psi _{k}^{b}\right|
\sigma _{k}\left| \psi _{k}^{b}\right\rangle .  \label{P_U^b}
\end{equation}
Thus, if Alice submits $\rho $ and realizes the convex decompositions $%
\left\{ \left( q_{k}^{0},\sigma _{k}^{0}\right) \right\} _{k=1}^{n_{0}}$ and
$\left\{ \left( q_{k}^{1},\sigma _{k}^{1}\right) \right\}
_{k=1}^{n_{1}}$ to unveil bit values of $0$ and $1$ respectively,
then if she is equally likely to wish to unveil $0$ as $1$ (as we
are assuming in this paper), her probability of unveiling the bit
of her choosing is
\begin{eqnarray}
P_{U} &=&\frac{1}{2}P_{U0}+\frac{1}{2}P_{U1}  \nonumber \\
&=&\frac{1}{2}\sum_{b=0}^{1}\sum_{k=1}^{n_{b}}q_{k}^{b}\left\langle
\psi _{k}^{b}\right| \sigma _{k}^{b}\left| \psi
_{k}^{b}\right\rangle .
\label{P_U}
\end{eqnarray}
The task is to maximize $P_{U}$ with respect to variations in $\rho ,$ $%
\left\{ \left( q_{k}^{0},\sigma _{k}^{0}\right) \right\} _{k=1}^{n_{0}}$ and
$\left\{ \left( q_{k}^{1},\sigma _{k}^{1}\right) \right\}
_{k=1}^{n_{0}}$ subject to the constraint that $\rho
=\sum_{k=1}^{n_{0}}q_{k}^{0}\sigma
_{k}^{0}=\sum_{k=1}^{n_{1}}q_{k}^{1}\sigma _{k}^{1}.$

It is useful to divide this optimization problem into two steps.
In the
first step one determines, for an arbitrary but fixed $\rho ,$ the $n_{b}$%
-element convex decomposition of $\rho $ that maximizes the probability $%
P_{Ub}$ of Alice unveiling the bit $b.$ Given this solution, the
probability $P_{U}$ of Alice unveiling the bit of her choosing
can be expressed entirely
in terms of the submitted $\rho .$ In the second step one determines the $%
\rho $ that maximizes $P_{U}.$

\section{Results for general protocols}
\label{V}

\subsection{\protect  The connection to state estimation}
\label{VA}
\noindent
We will show that the problem of optimizing the choice of convex
decomposition for an arbitrary but fixed density operator has an
intimate connection to the problem of optimal state estimation.
As discussed in section~3.2, for every convex decomposition
$\left\{ \left( q_{k},\sigma _{k}\right) \right\}$ there
exists a POVM $\left\{ E_{k}\right\} ,$ defined over the support
of $\rho ,$ that generates this decomposition as in Eq.~$\left(
\ref {connection}\right) $. Thus, Eq.~(\ref{P_U^b}) can be written
as
\[
P_{Ub}=\sum_{k=1}^{n_{b}}\left\langle \psi _{k}^{b}\right| \sqrt{\rho }E_{k}%
\sqrt{\rho }\left| \psi _{k}^{b}\right\rangle .
\]
A set of normalized states $\{\chi _{k}^{b}\}$ and probabilities $%
\{w_{k}^{b}\}$ can be defined in terms of $\rho $ and $\{\psi _{k}^{b}\}$ as
follows:
\begin{eqnarray}
\left| \chi _{k}^{b}\right\rangle  &=&\frac{\sqrt{\rho }\left| \psi
_{k}^{b}\right\rangle }{\sqrt{\left\langle \psi _{k}^{b}\right|
\rho \left|
\psi _{k}^{b}\right\rangle }},  \label{phi_k} \\
w_{k}^{b} &=&\frac{\left\langle \psi _{k}^{b}\right| \rho \left|
\psi _{k}^{b}\right\rangle }{\sum_{k}\left\langle \psi
_{k}^{b}\right| \rho
\left| \psi _{k}^{b}\right\rangle }.  \label{w_k}
\end{eqnarray}
In terms of these, $P_{Ub}$ has the form
\[
P_{Ub}=C\sum_{k=1}^{n_{b}}w_{k}^{b}\left\langle \chi
_{k}^{b}\right| E_{k}\left| \chi _{k}^{b}\right\rangle ,
\]
where $C=\sum_{k}\left\langle \psi _{k}^{b}\right| \rho \left|
\psi _{k}^{b}\right\rangle .$

We now recall~\cite{Helstrom} the problem of estimating the state
of a system that is known to have been prepared in one of $n_{b}$
states $\{\chi _{k}^{b}\}$ with prior probabilities
$\{w_{k}^{b}\}$. The most general type of measurement is a POVM
measurement, and it suffices to consider POVMs that have $n_{b}$
elements (this is established by an argument exactly analogous to
the one provided above for the sufficiency of $n_{b}$-element
decompositions in optmizing over coherent attacks). For a
measurement of the
POVM $\{E_{k}\},$ the probability of estimating correctly is $%
\sum_{k=1}^{n_{b}}w_{k}^{b}\left\langle \chi _{k}^{b}\right| E_{k}\left| \chi
_{k}^{b}\right\rangle .$

The connection between our problem and the state estimation
problem is now
clear. If $\{\chi _{k}^{b}\}$ and $\{w_{k}^{b}\}$ are defined by Eqs.~$%
\left( \ref{phi_k}\right) $ and $\left( \ref{w_k}\right) $, and $\{E_{k}\}$
is defined by $\left( \ref{connection}\right) $, then the
following relation holds. The probability of unveiling a bit $b$,
associated with a set of states $\left\{ \psi _{k}^{b}\right\} ,$
when Bob's reduced density operator
is $\rho ,$ given that Alice's strategy consists of realizing an $n_{b}$%
-element convex decomposition $\{(q_{k},\sigma _{k})\}$ of $\rho
,$ is a constant multiple of the probability of correctly
estimating the state of a system, known to be prepared in one of
$n_{b}$ states $\{\chi _{k}^{b}\}$
with prior probabilities $\{w_{k}^{b}\},$ given a measurement of the POVM $%
\{E_{k}\}.$

So, if one has the solution to the problem of finding the POVM
that maximizes the probability of correctly estimating the state
of a system from among a set of pure states, then one also has
the solution to the problem of finding the convex decomposition
of $\rho $ that Alice should realize to maximize her probability
of passing Bob's test. There is a {\em duality} between these two
information theoretic tasks.

This result is very useful since it connects a task about which
very little is known to one about which a great deal is known. In
particular, one is able to infer some general features of the
optimal cheat strategy by appealing to some well-known theorems
on state estimation.

One such feature is that if the $\left\{ \psi _{k}^{b}\right\} $
are linearly independent, and the support of $\rho $ is the span
of the $\left\{
\psi _{k}^{b}\right\} ,$ then the optimal convex decomposition of $\rho $ is an extremal decomposition. The proof is as follows. If the $\left\{ \psi
_{k}^{b}\right\} $ are linearly independent and span the support
of $\rho $, then the $\left\{
\chi _{k}^{b}\right\} $ are linearly independent. It is well known that in
estimating a state drawn from a set of linearly independent
states, the optimal POVM has elements of rank 1~\cite{Helstrom}.
The convex decomposition
that is associated with such a POVM has elements that are pure states, i.e., it is extremal.

\subsection{The support of the optimal density operator}
\label{VB}
\noindent
  We now turn to the problem of determining the optimal density
operator for Alice to submit to Bob. We begin by showing that
although Alice could cheat by submitting a system with more
degrees of freedom than the honest protocol specifies, she gains
no advantage by doing so. In other words, the optimal $\rho $ has
a support that is equal to or a subspace of the span of $\left\{
\psi _{k}^{0}\right\} _{k=1}^{n_{0}}\cup \left\{ \psi
_{k}^{1}\right\} _{k=1}^{n_{1}}.$ We establish this by
showing{\em \ }that for any $\rho ^{*}$ that has support strictly
greater than this span, there is a $\rho $ that has support that
is equal to or a subspace of this span and that yields a greater
value of $P_{U}.$ Suppose the optimal convex decomposition of
$\rho ^{*}$ for unveiling bit $b$ is denoted $\left\{
\left( q_{k}^{b*},\sigma _{k}^{b*}\right) \right\} _{k=1}^{n_{b}}.$ The
maximum probability of Alice unveiling the bit of her choosing
using $\rho ^{*}$ is then
\[
P_{U}^{\text{max}}\left( \rho ^{*}\right) =\frac{1}{2}\sum_{b=0}^{1}%
\sum_{k=1}^{n_{b}}\left\langle \psi _{k}^{b}\right| q_{k}^{b*}\sigma
_{k}^{b*}\left| \psi _{k}^{b}\right\rangle .
\]
However, if Alice submits the density operator
\[
\rho =G\rho ^{*}G/{\rm Tr}\left( \rho ^{*}G\right) ,
\]
where $G$ is the projector onto the span of $\left\{ \psi
_{k}^{0}\right\} _{k=1}^{n_{0}}\cup \left\{ \psi _{k}^{1}\right\}
_{k=1}^{n_{1}},$ and realizes the convex decomposition $\left\{
\left( q_{k}^{b},\sigma _{k}^{b}\right) \right\} _{k=1}^{n_{b}}$
defined by $q_{k}^{b}\sigma _{k}^{b}=Gq_{k}^{b*}\sigma
_{k}^{b*}G/{\rm Tr}\left( \rho ^{*}G\right) ,$ then her
probability of unveiling whatever bit she desires is
\[
P_{U}\left( \rho \right) =P_{U}^{\text{max}}\left( \rho ^{*}\right) /{\rm Tr}%
\left( \rho ^{*}G\right) .
\]
Since ${\rm Tr}\left( \rho ^{*}G\right) <1$, it follows that
$P_{U}\left(
\rho \right) >P_{U}^{\text{max}}\left( \rho ^{*}\right) .$

\subsection{Conditions for unveiling with certainty}
\label{VC}
\noindent
Finally, we consider the question of whether, for a particular
protocol, Alice can unveil the bit of her choosing with
certainty. The
necessary and sufficient condition for there to be a strategy that makes $%
P_{Ub}=1$ for a given $b,$ is that $\rho $ is decomposed by the
set of states $\left\{ \psi _{k}^{b}\right\} _{k=1}^{n_{b}}$,
that is, there must exist a probability distribution $\left\{
q_{k}^{b}\right\} _{k=1}^{n_{b}}$ such that $\left\{ \left(
q_{k}^{b},\left| \psi _{k}^{b}\right\rangle
\left\langle \psi _{k}^{b}\right| \right) \right\} _{k=1}^{n_{b}}$ forms a
convex decomposition of $\rho .$ The necessary and sufficient
condition for there to be a strategy that makes $P_{U}=1$ is that
there exists a $\rho $ that is decomposed by both $\left\{ \psi
_{k}^{0}\right\} _{k=1}^{n_{0}}$ and $\left\{ \psi
_{k}^{1}\right\} _{k=1}^{n_{1}}$. In the terminology of section~3.1, $\left\{ \psi _{k}^{0}\right\} _{k=1}^{n_{0}}$ and $\left\{
\psi _{k}^{1}\right\} _{k=1}^{n_{1}}$ must be composably coincident.

The results described in this section constitute all that we
shall say about the optimal cheat strategy for an arbitrary
protocol. For the rest of this paper, we shall restrict ourselves
to the special case of sets $\left\{ \psi _{k}^{0}\right\}
_{k=1}^{n_{0}}$ and $\left\{ \psi _{k}^{1}\right\}
_{k=1}^{n_{1}}$ whose union span at most a two dimensional
Hilbert space, that is, protocols that can be implemented using a
single qubit.

\section{Results for qubit protocols}
\label{VI}

\subsection{The Bloch ball representation}
\label{VIA}
\noindent
  Our optimization problem is greatly simplified in the case of a
2D Hilbert space since there is a one-to-one mapping between the
set of all density operators in such a space and the set of all
points within the unit ball of ${\Bbb R}^{3}$. For clarity, we
begin by reminding the reader about the details of this mapping.

If one defines an inner product between operators $A$ and $B$ by ${\rm Tr}%
\left( A^{\dag }B\right) ,$ the set of operators over a Hilbert space forms
an inner product space. In a 2d Hilbert space, a particularly
convenient
orthogonal basis for the set of operators is the set of Pauli operators $%
\left\{ \sigma _{x},\sigma _{y},\sigma _{z},I\right\} $, with matrix
representations in the $\left\{ \left| 0\right\rangle ,\left|
1\right\rangle
\right\} $ basis of
\begin{equation}
\sigma _{x} =\left(
\begin{array}{ll}
0 & 1 \\
1 & 0
\end{array}
\right) ,\text{ }\sigma _{y}=\left(
\begin{array}{ll}
0 & -i \\
i & 0
\end{array}
\right) ,\text{ } 
\sigma _{z} =\left(
\begin{array}{ll}
1 & 0 \\
0 & -1
\end{array}
\right) ,\text{ }I=\left(
\begin{array}{ll}
1 & 0 \\
0 & 1
\end{array}
\right) .
\end{equation}
Any operator $A$ can therefore be written as $A=\frac{1}{2}\left( a_{0}I+%
\vec{a}\cdot \vec{\sigma}\right) $ where $\vec{\sigma}{\bf =}\left( \sigma
_{x},\sigma _{y},\sigma _{z}\right) $ and $\vec{a}=\left(
a_{x},a_{y},a_{z}\right) ,$ with $a_{0},a_{x},a_{y},a_{z}\in
{\Bbb C}^{1}.$ In particular, for a density operator $\rho ,$ the
constraints of unit trace (${\rm Tr}(\rho )=1)$ and positivity
$(\det \left( \rho \right) \ge 0)$ imply that
\begin{eqnarray*}
\rho  &=&\frac{1}{2}\left( I+\vec{r}\cdot \vec{\sigma}\right) , \\
\mbox{where }\vec{r} &\in &{\Bbb R}^{3}\mbox{ and }\left| \vec{r}\right| \le
1.
\end{eqnarray*}
Thus we see that every density operator is represented by a
vector $\vec{r}$ within the unit ball of ${\Bbb R}^{3}$, which we
shall refer to as the Bloch ball.\fnm{e}\fnt{e}{The surface of
the ball is usually referred to as the Bloch sphere or the
Riemann sphere in the context of spins, and the Poincare sphere
in the context of photon polarization.}

Density operators describing pure states are characterized by a
vanishing determinant which
corresponds to a vector of
unit length, $\left| \vec{r}\right| =1$, which we shall sometimes denote by $\hat{r}$. Thus, pure states are represented by the points on the surface of
the ball. The completely mixed state, $\rho =\frac{1}{2}I,$ is
represented by $\vec{r}=\vec{0}{\bf ,}$ which is the point at the
centre of the ball. If two density operators $\rho _{1}$ and
$\rho _{2}$ are represented by vectors
$\vec{r}_{1}$ and $\vec{r}_{2},$ the inner product between $\rho _{1}$ and $%
\rho _{2}$ is given by ${\rm Tr}\left( \rho _{1}\rho _{2}\right) =\frac{1}{2}%
\left( 1+\vec{r}_{1}\cdot \vec{r}_{2}\right) .$ It follows that orthogonal
states are represented by antipodal points, since ${\rm Tr}(\rho
_{1}\rho _{2})=0$ implies $\vec{r}_{1}\cdot \vec{r}_{2}=-1.$

We are now in a position to obtain a representation on the Bloch
ball of all the density operators that can be formed by convex
combination of a particular set of elements $\left\{ \sigma
_{k}\right\} _{k=1}^{n},$ that is, all the $\rho $ that have the
form
\[
\rho =\sum_{k=1}^{n}q_{k}\sigma _{k}
\]
for some probability distribution $\{q_{k}\}_{k=1}^{n}.$ Since
the set of
density operators that can be formed by an arbitrary set of elements $%
\left\{ \sigma _{k}\right\} _{k=1}^{n}$ is the same as the set that can be
formed by an uncontractable set of elements from which all the states in $%
\left\{ \sigma _{k}\right\} _{k=1}^{n}$ can be built up by convex
combination, it suffices to consider only uncontractable sets of
elements.

Denoting the Bloch vectors associated with $\rho $ and $\sigma _{k}$ by $%
\vec{r}$ and $\vec{s}_{k}$ respectively, it is easy to see that the relevant set is
given by
\[
\vec{r}=\sum_{k=1}^{n}q_{k}\vec{s}_{k},\mbox{ where }0\le q_{k}\le 1,\text{ }%
\sum_{k=1}^{n}q_{k}=1.
\]
To understand what this manifold of points looks like, consider
the simplest case of $n=2$. The above equation can then be
written as
\[
\vec{r}=\vec{s}_{1}+\lambda \left( \vec{s}_{2}-\vec{s}_{1}\right) ,\mbox{
where }0\le \lambda \le 1.
\]
This is simply the parametric equation for a segment of a
straight line extending between $\vec{s}_{1}$ and $\vec{s}_{2}.$
Similarly, in the case of $n=3,$ we have
\[
\vec{r}=\vec{s}_{1}+\lambda \left( \vec{s}_{2}-\vec{s}_{1}\right) +\xi
\left( \vec{s}_{3}-\vec{s}_{2}\right),
\]
where $0\le \lambda \le 1$ and $0\le \xi \le \lambda.$
Since $\sigma _{1},\sigma _{2}$ and $\sigma _{3}$ were assumed to
form an
uncontractable set of elements, $\vec{s}_{1}$, $\vec{s}_{2}$ and $\vec{s}%
_{3} $ cannot lie on a line, and therefore define the vertices of
a triangle. The above equation is the parametric equation for the
surface of points inside this triangle$.$ For $n=4,$ we have
\[
\vec{r} =\vec{s}_{1}+\lambda \left( \vec{s}_{2}-\vec{s}_{1}\right) +\xi
\left( \vec{s}_{3}-\vec{s}_{2}\right) +\zeta \left( \vec{s}_{4}-\vec{s}%
_{3}\right),
\]
where $0 \le \lambda \le 1$, $0\le \xi \le \lambda$ and %
$0\le \zeta \le \xi$.
Again, since $\sigma _{1},\sigma _{2}$, $\sigma _{3}$ and $\sigma
_{4}$ were
assumed to form an uncontractable set of elements, none of $\vec{s}_{1}$, $%
\vec{s}_{2}$, $\vec{s}_{3}$ or $\vec{s}_{4}$ can lie along the line segment
defined by any other two, nor inside the surface of the triangle
defined by any other three, and thus these vectors define the
vertices of either a convex quadrilateral or a tetrahedron. The
above equation is the parametric equation for the surface of
points inside this quadrilateral, or the volume of points inside
this tetrahedron. Similarly, for greater than $4$ uncontractable
elements, we obtain the parametric equation for the points inside
an $n$-vertex convex polygon or convex polyhedron. All told, in
the case of a set of $n$ uncontractable elements, the set of
density operators
that can be composed from these will be represented by the region inside an $%
n$-vertex convex polytope. A few different sets of states and the
density operators that can be composed from them are depicted in
Fig.~1.

\begin{figure}[htbp]

\centerline{\epsfig{file=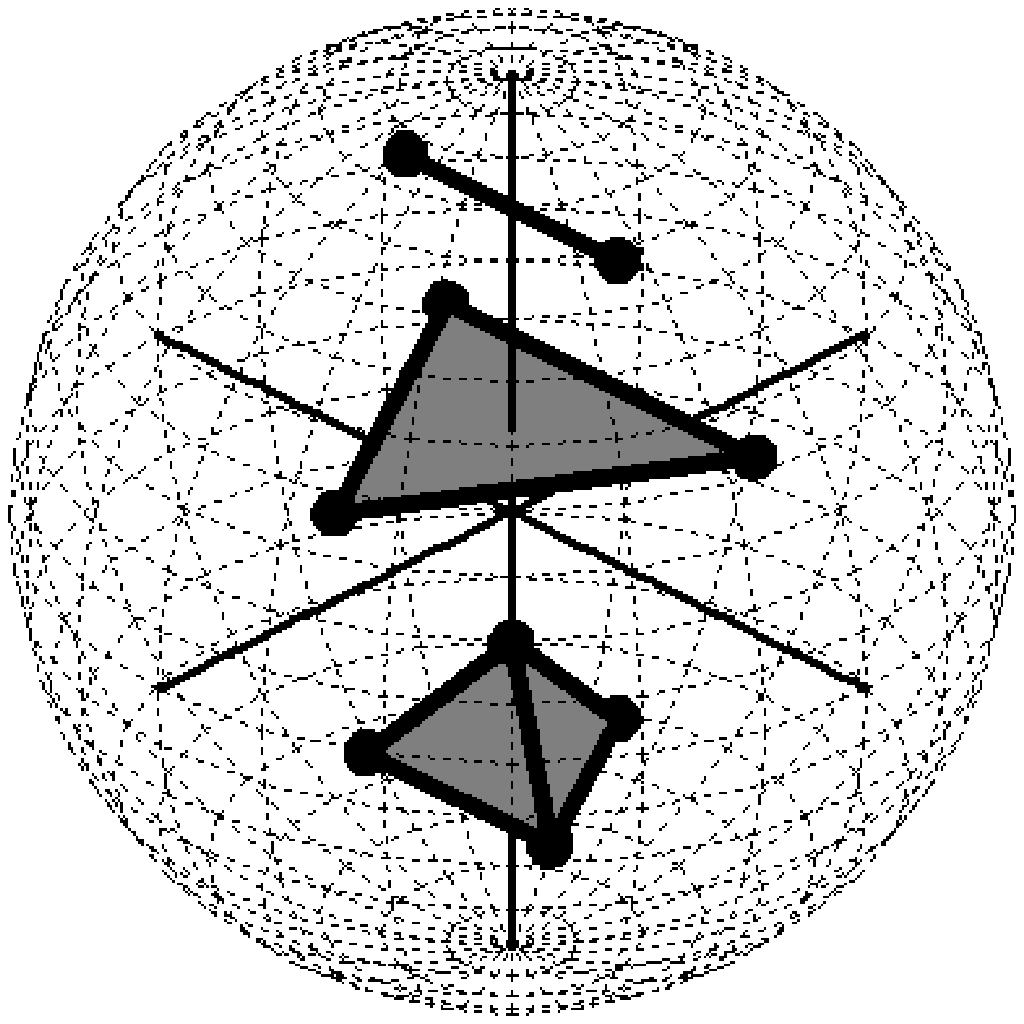,width=50mm,bbllx=5cm,bblly=9cm,bburx=18cm,bbury=21.5cm,clip=}}
{\fcaption{ A depiction of three sets of states containing 2, 3
and 4 elements respectively. The points in the Bloch ball
representing these states are indicated by small black spheres.
The manifolds inside the line segment, triangle and tetrahedron
that are defined by each set of points represent all the density
operators that can be composed with each set of states.}}
\end{figure}

It is now easy to see the solution to a complementary problem,
namely, how to obtain a representation on the Bloch ball of all
the uncontractable convex decompositions of a particular density
operator $\rho $. If $\rho $ is represented by the point
$\vec{r}{\bf ,}$ then every $n$-element
uncontractable convex decomposition of $\rho $ is represented by an $n$%
-vertex convex polytope which contains $\vec{r}$. For instance,
every $2$-element uncontractable convex decomposition of $\rho $
is represented by a line segment that contains $\vec{r}{\bf ;}$
every $3$-element uncontractable
convex decomposition of $\rho $ is represented by a triangle that contains $%
\vec{r}{\bf ,}$ and so forth. In Fig.~2, we illustrate a few of the convex
decompositions of a fixed density operator.

\begin{figure}[tbp]

\centerline{\epsfig{file=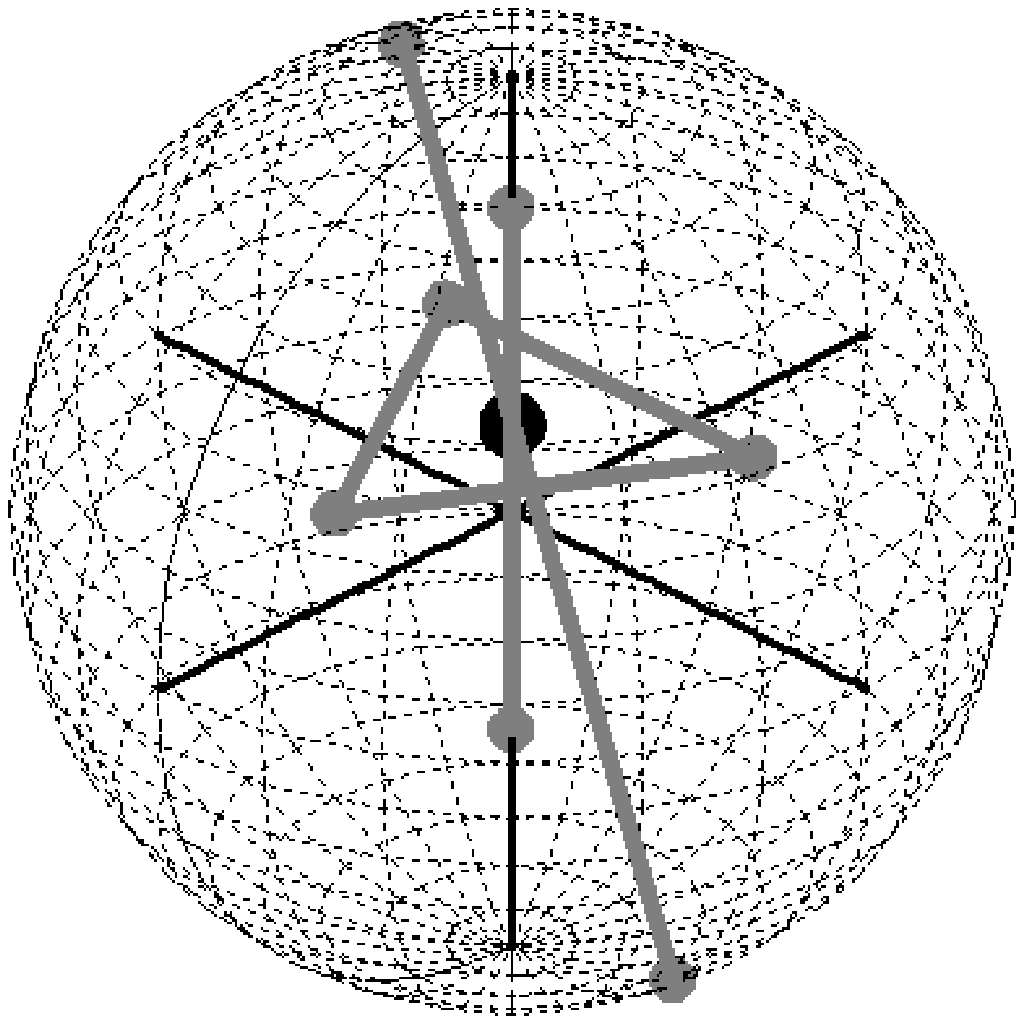,width=50mm,bbllx=5cm,bblly=9cm,bburx=18cm,bbury=21.5cm,clip=}}
{\fcaption{An illustration of three convex decompositions of a
fixed density operator, two of which are 2-element
decompositions, and one of which is a 3-element decomposition.
The point in the Bloch ball representing the density operator is
indicated with a large black sphere. Each convex decomposition is
represented by a polytope containing the point representing the
density operator, the vertices of which represent the elements of
the decomposition. These are indicated in grey. The longer of the
two line segments, which has its vertices on the surface of the
Bloch ball, is an example of an extremal convex decomposition.}}
\end{figure}

Of particular interest to us in the present context are {\em
extremal} convex decompositions of a density operator $\rho $
(which are always uncontractable). Since pure states are
associated with unit Bloch vectors, the convex polytopes
associated with such decompositions have their vertices on the
surface of the Bloch ball. Fig.~2 provides an example of this
distinction.

\vfill\eject
\subsection{The conditions under which Alice can unveil the bit of her
choosing with certainty}
\label{VIB}
\noindent
In section~5.3 it was pointed out that a strategy with $P_{Ub}=1$
exists if and only if $\rho $ is decomposed by the states
$\left\{ \psi _{k}^{b}\right\} .$ The Bloch ball
representation gives a simple way of testing whether this
condition is satisfied for protocols restricted to a 2D Hilbert
space. It suffices to plot the convex polytope whose vertices are
the points representing the $\left\{ \psi _{k}^{b}\right\}
$ and to determine whether the point representing $\rho
$ is contained in this polytope or not. If it is, then Alice can
unveil bit $b$ with certainty. If it is not, then she cannot. An
example of the two possibilities is provided in Fig.~3.

\begin{figure}[hbp]

\centerline{\epsfig{file=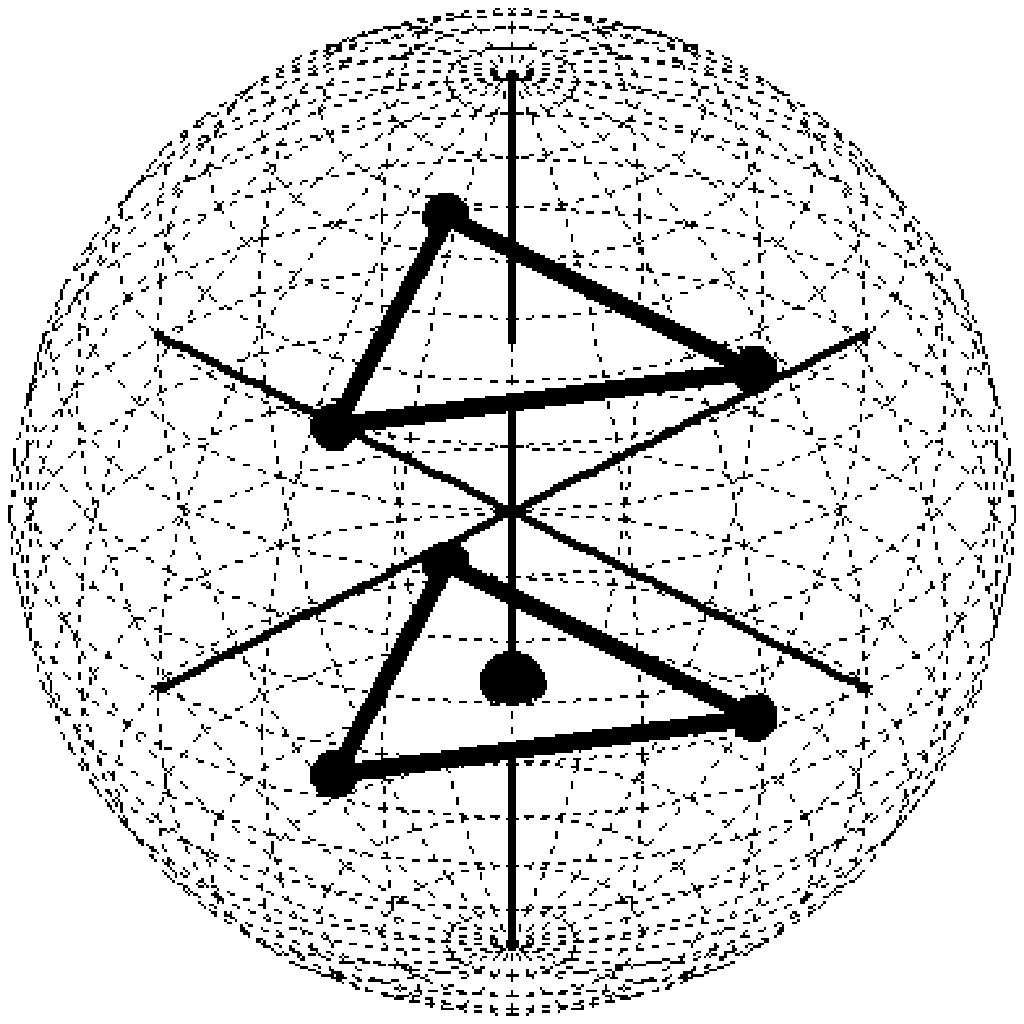,width=50mm,bbllx=5cm,bblly=9cm,bburx=18cm,bbury=21.5cm,clip=}}
\fcaption{ A depiction of a fixed density operator and two
sets of states, the lower of which decomposes the density
operator and the uppermost of which does not. }
\end{figure}

More importantly, we can now answer the question of whether there
exists a strategy for Alice with $P_{U}=1$ for protocols
restricted to a 2D Hilbert space. As pointed out in section~5.3
this only occurs if the two sets of states are composably
coincident. It is clear now how to verify whether this is the
case or not. Simply plot the convex polytopes associated with both
sets of states, and determine whether they intersect one another
or not. If they do, then any point inside the region of
intersection corresponds to a density operator that is decomposed
by both sets and consequently lets Alice unveil the bit of her
choosing with probability $1$. If they do not, then this
probability is strictly less than $1.$

The convex polytopes associated with the sets of states used in
the BB84 BC protocol are depicted in Fig.~4. Since these cross at
the origin, it follows
that if Alice submits to Bob the completely mixed state, she can achieve $%
P_{U}=1.$ So we simply have a restatement of the fact that if
Alice initially prepares a maximally entangled state, such as the
EPR state, and submits half to Bob, then she can achieve
$P_{U}=1.$ The protocols we shall consider in the rest of this
paper are associated with non-intersecting convex polytopes. See
Figs. 6-10 for examples.

\begin{figure}[htbp]
\centerline{\epsfig{file=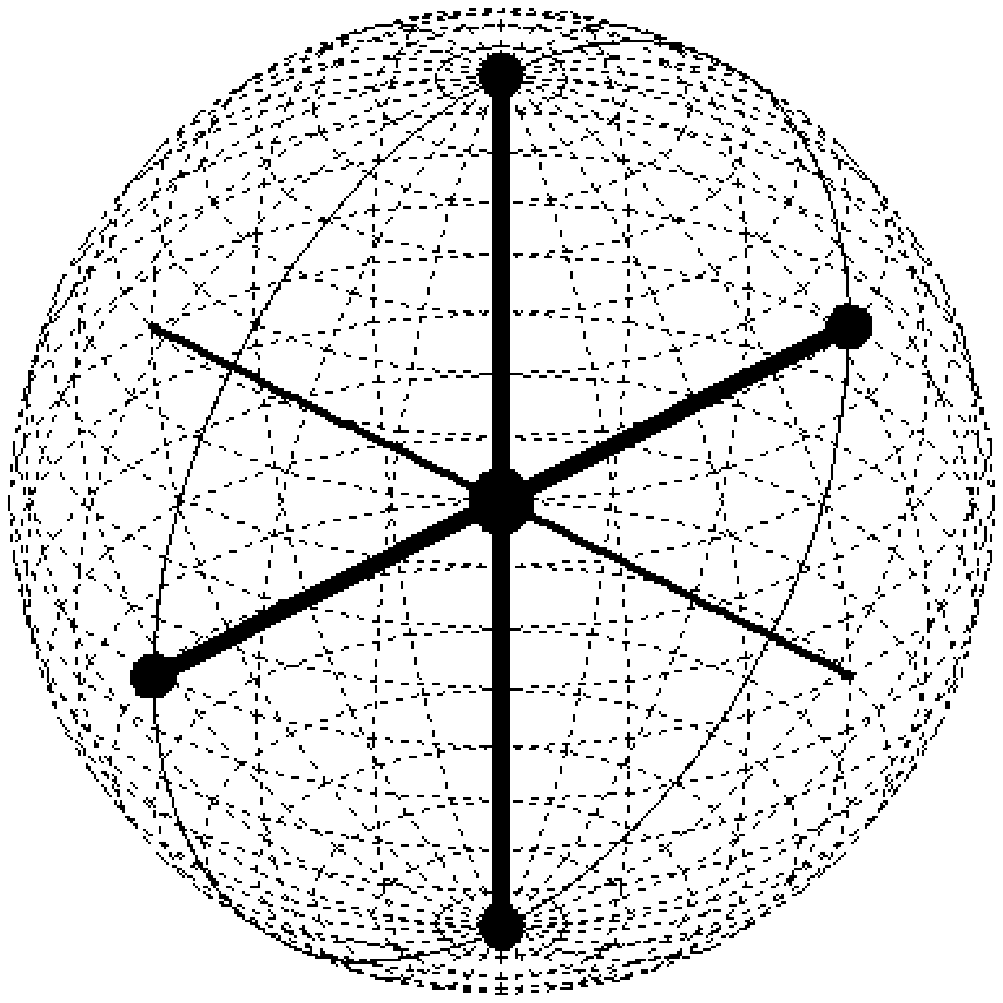,width=50mm,bbllx=5cm,bblly=9cm,bburx=18cm,bbury=21.5cm,clip=}}

\fcaption{ The Bloch ball representation of the BB84 BC
protocol. Since the polytopes representing the sets of states
defined by the protocol intersect, Alice can submit the density
operator associated with their intersection to make her
probability of unveiling whatever bit she desires equal to unity.
}
\end{figure}

\subsection{Optimizing over the convex decompositions of an arbitrary but
fixed density operator}
\label{VIC}
\noindent
We now turn to the problem of determining the optimal EPR
cheating strategy for a qubit protocol. We do not solve this
problem completely; rather, we solve it under the further
restriction that each set contains only linearly independent
states. In the present 2D context, linear independence implies
that each set can have no more than two elements.

As discussed in section~4, it is useful to split the problem
into two parts involving optimization over convex decompositions
of an arbitrary but fixed density operator, followed by
optimization over density operators. We address these two parts
of the problem in this section and the next section respectively.

We begin with the problem of maximizing the probability,
$P_{Ub},$ that
Alice can unveil the bit $b$ given that she submitted a density operator $%
\rho .$ This maximum must be found with respect to variations in the convex
decomposition of $\rho $ that she realizes. The optimal
decomposition will depend on $\rho $ and the states in the set
$\left\{ \psi _{k}^{b}\right\} $. In order to
simplify the notation in this section, we drop the index $b$ from
$\left| \psi _{k}^{b}\right\rangle $ and $n_{b}$. We also assume
that $\rho $ is impure, since otherwise there is no optimization
problem to be solved.

\subsubsection{A set containing one element $(n=1)$}
\label{VIC1}

In this case, Bob's test is fixed (he always measures the projector onto $%
\left| \psi _{1}\right\rangle $), so the probability of passing this test
depends only on $\rho $ and not on the convex decomposition of
$\rho $ that Alice realizes. Thus, there is no optimization over
decompositions to be performed in this case.

\subsubsection{A set containing two elements $(n=2)$}
\label{VIC2}

Let the Bloch vectors associated with $\left| \psi _{k}\right\rangle $ and $%
\rho $ be denoted by $\hat{a}_{k}$ and $\vec{r}$ respectively, and let those
associated with the elements, $\sigma _{k},$ of the two-element
convex decomposition $\left\{ \left( q_{k},\sigma _{k}\right)
\right\} _{k=1}^{2}$ that Alice realizes be denoted by
$\vec{s}_{k}.$ In terms of these, Alice's probability of passing
Bob's test, specified by Eq.~(\ref{P_U^b}), has the following form
\begin{equation}
P_{Ub}=\frac{1}{2}\left( 1+\sum_{k=1}^{2}q_{k}\left( \hat{a}_{k}\cdot \vec{s}%
_{k}\right) \right) .  \label{P_Ub Bloch}
\end{equation}
We must maximize this subject to the constraint that $\vec{r}=\sum_{k}q_{k}%
\vec{s}_{k}.$

We find that the optimal convex decomposition of $\vec{r}$ is
given by
\begin{eqnarray}
\vec{s}_{1}^{\;\text{opt}} &=&\vec{r}+L_{+}\left( \vec{r}\right) \hat{d},
\nonumber \\
\vec{s}_{2}^{\;\text{opt}} &=&\vec{r}+L_{-}\left( \vec{r}\right) \hat{d},
\label{optdecomp4fixedrho}
\end{eqnarray}
and
\begin{equation}
q_{k}^{\text{opt}}=\frac{1}{2}\frac{1-\left| \vec{r}\right| ^{2}}{1-\vec{r}%
\cdot \vec{s}_{k}^{\;\text{opt}}}
\end{equation}
where
\begin{equation}
L_{\pm }\left( \vec{r}\right) =-\vec{r}\cdot \hat{d}\pm \sqrt{1-\left| \vec{r%
}\right| ^{2}+\left( \vec{r}\cdot \hat{d}\right) ^{2}},
\label{Lpm}
\end{equation}
and
\begin{equation}
\hat{d}=\frac{\hat{a}_{1}-\hat{a}_{2}}{\left| \hat{a}_{1}-\hat{a}_{2}\right|
}.  \label{d}
\end{equation}
Note that $\left| \vec{s}_{1}^{\;\text{opt}}\right| =\left| \vec{s}_{2}^{\;\text{%
opt}}\right| =1$, which means that this is an extremal convex
decomposition. The proof of optimality is presented in Appendix A.

This solution has a very simple geometrical description. It is
the convex decomposition that is represented by the chord (line segment whose endpoints lie on the surface of the
ball) that contains $\vec{r}$ and that is parallel to the chord
defined by $\hat{a}_{1}$, $\hat{a}_{2}.$
  An example is presented in Fig.~5.

\begin{figure}[htbp]
\centerline{\epsfig{file=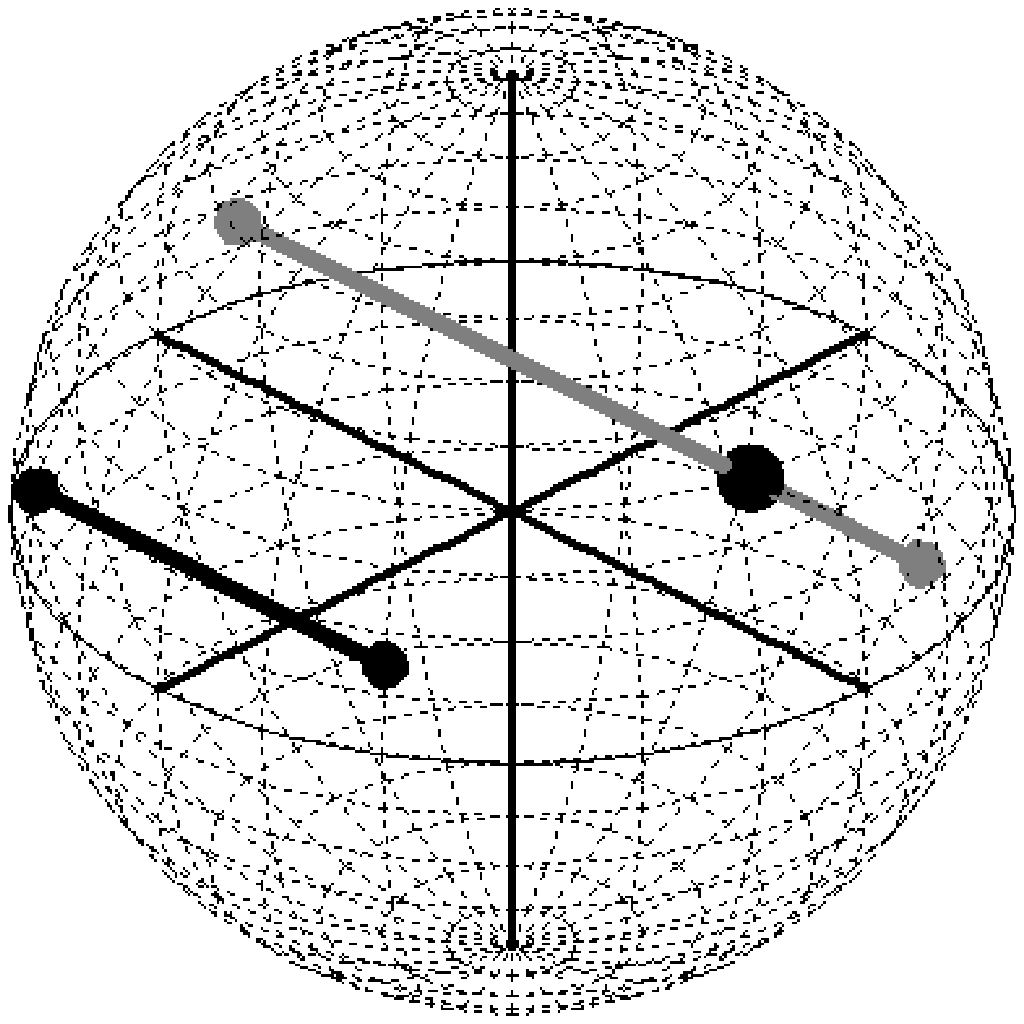,width=50mm,bbllx=5cm,bblly=9cm,bburx=18cm,bbury=21.5cm,clip=}}
\fcaption{ An illustration of the optimal convex
decomposition for Alice to realize when she has submitted to Bob
a fixed density operator(indicated by the large black sphere) and
is attempting to convince him that he has one of two states
(indicated by the small black spheres). This is represented by the
chord(indicated in grey) that is parallel to the chord defined by
the two states. After Alice realizes this decomposition (by
making a measurement on the system that is entangled with Bob's),
she updates her description of Bob's system to whichever of the
elements it happened to be collapsed to(indicated by the grey
spheres). When it comes time for Alice to announce to Bob which
of the two states he should test for to verify her honesty, she
announces the state which has the smallest angular separation
from the element of the decomposition onto which she has
collapsed his system.}
\end{figure}

The corresponding probability of passing Bob's test is simply
\[
P_{Ub}^{\text{max}}=\frac{1}{2}\left( 1+\left( \vec{r}+L_{+}\left( \vec{r}%
\right) \hat{d}\right) \cdot \hat{a}_{1}\right) .
\]
In Hilbert space language,
\begin{eqnarray*}
&&P_{Ub}^{\text{max}}=\frac{1}{2}\left( \left\langle \psi
_{1}|\rho |\psi _{1}\right\rangle +\left\langle \psi _{2}|\rho
|\psi _{2}\right\rangle
\right)  \\
&&+\sqrt{2\left( 1-{\rm Tr}(\rho ^{2})\right) \left| \left\langle
\psi _{1}|\psi _{2}\right\rangle \right| ^{2}+\left( \left\langle
\psi _{1}|\rho |\psi _{1}\right\rangle -\left\langle \psi
_{2}|\rho |\psi _{2}\right\rangle
\right) ^{2}}.
\end{eqnarray*}

\subsection{\protect  Optimizing over density operators}
\label{VID}
\noindent
We now consider the problem of determining the optimal density operator for Alice to submit to Bob in order to maximize her probability of
unveiling the bit of her choosing. The solution will depend on
the values of
$n_{0}$ and $n_{1}.$ Given that we are assuming that the states in the sets $%
\left\{ \psi _{k}^{0}\right\} _{k=1}^{n_{0}}$ and $\left\{ \psi
_{k}^{1}\right\} _{k=1}^{n_{1}}$ are linearly independent, there
are only three possibilities to address: both sets contain two
elements; one set contains two elements and the other contains
one element; both sets contain one element. We shall consider
each of these in turn.

\subsubsection{Both sets contain two elements $(n_{0}=n_{1}=2)$}
\label{VID1}

Denote the Bloch vector associated with the state $\left| \psi
_{k}^{b}\right\rangle $ by $\hat{a}_{k}^{b}.$ The result of the
previous section indicates that whatever the optimal $\vec{r}$
is, the optimal convex decomposition for unveiling bit $b$ is
represented by the chord passing
through $\vec{r}$ parallel to the chord defined by $\hat{a}_{1}^{b},\hat{a}%
_{2}^{b}.$ We therefore have that Alice's probability of
unveiling the bit of her choosing given an arbitrary $\vec{r}$
and given that when she
attempts to unveil the bit $b$ she realizes the convex decomposition of $%
\vec{r}$ that is optimal for doing so, is simply
\begin{equation}
P_{U}=\sum_{b=0}^{1}\frac{1}{4}\left( 1+\left( \vec{r}+L_{b+}\left( \vec{r}%
\right) \hat{d}_{b}\right) \cdot \hat{a}_{1}^{b}\right) ,  \label{P_U Bloch}
\end{equation}
where $\hat{d}_{b}=\frac{\hat{a}_{1}^{b}-\hat{a}_{2}^{b}}{\left| \hat{a}%
_{1}^{b}-\hat{a}_{2}^{b}\right| }$ and
\[
L_{b+}\left( \vec{r}\right) =-\vec{r}\cdot \hat{d}_{b}+\sqrt{1-\left| \vec{r}%
\right| ^{2}+\left( \vec{r}\cdot \hat{d}_{b}\right) ^{2}}.
\]

It will be convenient to adopt the convention that the states in
the protocol are indexed in such a way that $\left\langle \psi
_{1}^{0}|\psi _{1}^{1}\right\rangle =\max_{k,k^{\prime
}}\left\langle \psi _{k}^{0}|\psi _{k^{\prime }}^{1}\right\rangle
.$ In terms of the Bloch ball, the
convention states that if one draws the chords defined by $\hat{a}_{1}^{0},%
\hat{a}_{2}^{0}$ and $\hat{a}_{1}^{1},\hat{a}_{2}^{1},$ the endpoints $\hat{a%
}_{1}^{0}$ and $\hat{a}_{1}^{1}$ have the smallest separation.

We consider two cases.

  {\bf Case 1: }The chords defined by $\hat{a}_{1}^{0},\hat{a}%
_{2}^{0}$ and $\hat{a}_{1}^{1},\hat{a}_{2}^{1}$ are parallel.

  In this case $\hat{d}_{0}=\hat{d}_{1}$ and there are a family of
optimal $\vec{r}$'s satisfying the parametric equation
\begin{equation}
\vec{r}^{\;\text{opt}}=\frac{\hat{a}_{1}^{0}+\hat{a}_{1}^{1}}{\left| \hat{a}%
_{1}^{0}+\hat{a}_{1}^{1}\right| }+\lambda \hat{d}_{0},\mbox{ for
}0\le
\lambda \le 2\frac{\hat{a}_{1}^{0}+\hat{a}_{1}^{1}}{\left| \hat{a}_{1}^{0}+%
\hat{a}_{1}^{1}\right| }\cdot \hat{d}_{0}.  \label{ropt case1}
\end{equation}
This family corresponds to the points on the chord of the Bloch
ball that is
parallel to the chord defined by $\hat{a}_{1}^{0},\hat{a}_{2}^{0}$ (or $%
\hat{a}_{1}^{1},\hat{a}_{2}^{1})$ and that passes through the point on the
surface of the ball that is equidistant between $\hat{a}_{1}^{0}$ and $%
\hat{a}_{1}^{1}$. This is illustrated in Fig.~7.

{\bf Case 2:} The chords defined by $\hat{a}_{1}^{0},\hat{a}_{2}^{0}$ and $%
\hat{a}_{1}^{1},\hat{a}_{2}^{1}$ are not parallel.

In this case, the optimal $\vec{r}$ is unique and is given by
\begin{equation}
\vec{r}^{\;\text{opt}}=\left\{
\begin{array}{r}
\vec{r}^{\;\text{max}}\mbox{ if }\left| \vec{r}^{\;\text{max}}\right| \le 1 \\
\frac{\hat{a}_{1}^{0}+\hat{a}_{1}^{1}}{\left| \hat{a}_{1}^{0}+\hat{a}%
_{1}^{1}\right| }\mbox{ otherwise}
\end{array}
\right. ,  \label{r opt}
\end{equation}
where
\begin{equation}
\vec{r}^{\;\text{max}}=x_{0}^{\text{max}}\hat{d}_{1}^{\perp }+x_{1}^{\text{max}%
}\hat{d}_{0}^{\perp }+x_{2}^{\text{max}}\hat{n}.  \label{r max}
\end{equation}
Here
\begin{eqnarray}
x_{0}^{\text{max}} &=&\frac{1}{\gamma _{1}}\hat{a}_{1}^{1}\cdot \hat{d}%
_{1}^{\perp }\sqrt{1-\left( x_{2}^{\text{max}}\right) ^{2}},  \nonumber \\
x_{1}^{\text{max}} &=&\frac{1}{\gamma _{0}}\hat{a}_{1}^{0}\cdot \hat{d}%
_{0}^{\perp }\sqrt{1-\left( x_{2}^{\text{max}}\right) ^{2}},  \nonumber \\
x_{2}^{\text{max}} &=&\frac{\left(
\hat{a}_{1}^{0}+\hat{a}_{1}^{1}\right)
\cdot \hat{n}}{\sqrt{\left( \left( \hat{a}_{1}^{0}+\hat{a}_{1}^{1}\right)
\cdot \hat{n}\right) ^{2}+(\gamma _{0}+\gamma _{1})^{2}}},  \label{x opt}
\end{eqnarray}
where $\gamma _{b}=\sqrt{1-\left( \hat{a}_{1}^{b}\cdot
\hat{n}\right) ^{2}}$ and where
\begin{eqnarray}
\hat{n} &=&\hat{d}_{0}\times \hat{d}_{1},  \nonumber \\
\hat{d}_{b}^{\perp } &=&\hat{d}_{b}\times \hat{n},  \nonumber \\
\hat{d}_{b} &=&\frac{\hat{a}_{1}^{b}-\hat{a}_{2}^{b}}{\left| \hat{a}_{1}^{b}-%
\hat{a}_{2}^{b}\right| }.  \label{nonorth Bloch basis}
\end{eqnarray}

Thus, the solution has one of two forms depending on whether the condition $%
\left| \vec{r}^{\;\text{max}}\right| \le 1$ holds or not. If it does not hold$,
$ then $\vec{r}^{\;\text{opt}}=\left(
\hat{a}_{1}^{0}+\hat{a}_{1}^{1}\right) /\left|
\hat{a}_{1}^{0}+\hat{a}_{1}^{1}\right| ,$ which is simply the
point on the surface of the Bloch ball that is equidistant
between $\hat{a}_{1}^{0} $ and $\hat{a}_{1}^{1}$ along the
geodesic which connects them (recall that in our labelling
convention $\hat{a}_{1}^{0}$ and $\hat{a}_{1}^{1}$ are the
closest endpoints of the chords defined by
$\hat{a}_{1}^{0},\hat{a}_{2}^{0}$ and
$\hat{a}_{1}^{1},\hat{a}_{2}^{1}$). Fig.~6 provides an example of
a BC
protocol where this is the case. If the condition $\left| \vec{r}^{\;\text{max}%
}\right| \le 1$ {\em does} hold, then $\vec{r}^{\;\text{opt}}=\vec{r}^{\text{%
max}}$. We will not attempt to provide a geometrical description
of this point in the general case, however Figs. 4 and 9 provide
simple examples of BC protocols where $\left|
\vec{r}^{\;\text{max}}\right| \le 1$.

\begin{figure}[htbp]
\centerline{\epsfig{file=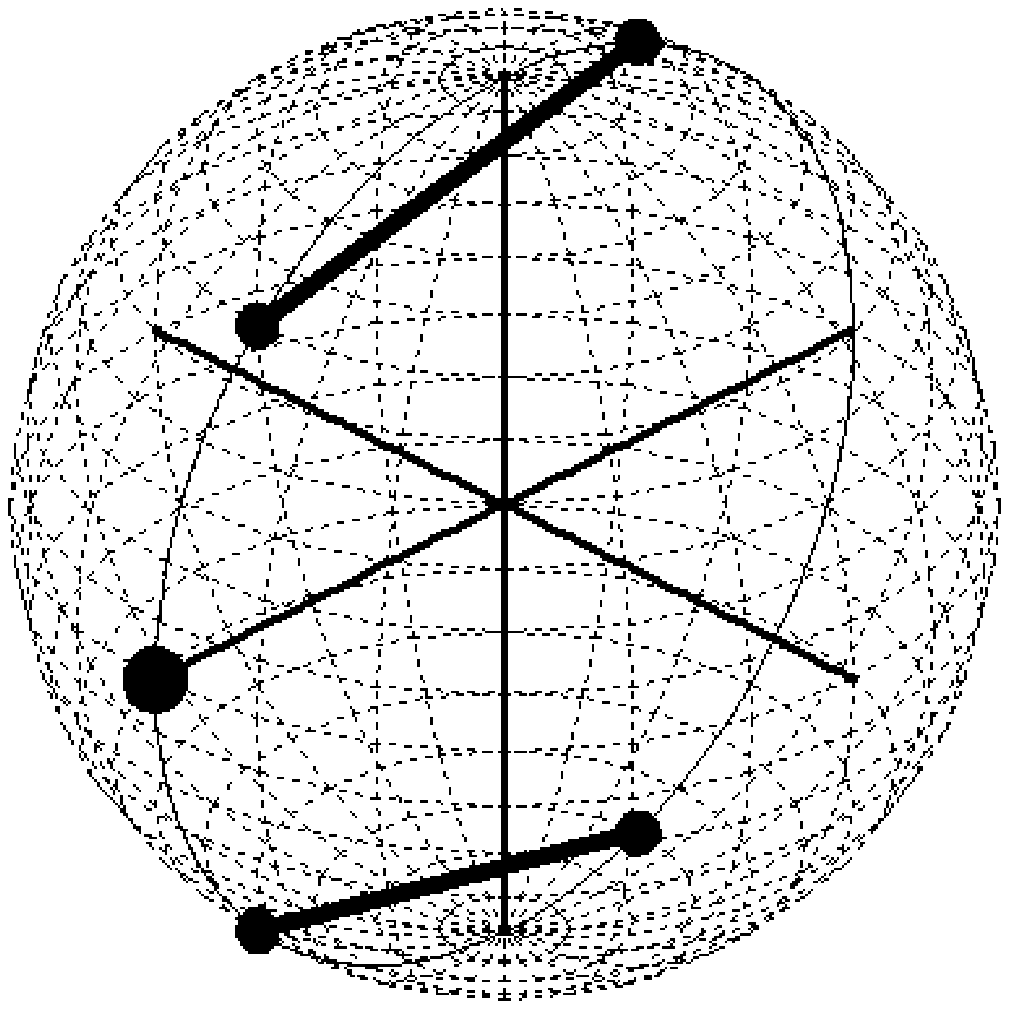,width=50mm,bbllx=5cm,bblly=9cm,bburx=18cm,bbury=21.5cm,clip=}}

\fcaption{ A BC protocol where the two sets of states are
represented by chords that lie in a plane, but which do not
intersect inside the Bloch ball. The optimal density operator is
represented by the point that lies equidistant between the two
closest chord endpoints on the geodesic which connects them.}
\end{figure}

In situations having a high degree of symmetry, one can easily
deduce some of the features of $\vec{r}^{\;\text{opt}}.$ We present
a few such cases.

{\bf Case 2.1: }If the chord defined by $\hat{a}_{1}^{0}$ and $\hat{a}%
_{2}^{0}$ and the chord defined by $\hat{a}_{1}^{1}$ and
$\hat{a}_{2}^{1}$ lie in a plane, then $\vec{r}^{\;\text{max}}$ is
the point of intersection of the lines containing these chords.
If this point falls inside the Bloch ball ($\left|
\vec{r}^{\;\text{max}}\right| \le 1$), then it represents the
optimal density operator. This confirms the results of section~5.3. The BB84 BC protocol, illustrated in Fig.~4, is an instance
of such a case. If
the point of intersection falls outside the Bloch ball ($\left| \vec{r}^{%
\;\text{max}}\right| >1$), then the optimal density operator is as described
above. The BC protocol that is illustrated in Fig.~6 is an
instance of such a case.

{\bf Case 2.2: }If the chord defined by $\hat{a}_{1}^{0}$ and $\hat{a}%
_{2}^{0}$ and the chord defined by $\hat{a}_{1}^{1}$ and
$\hat{a}_{2}^{1}$ both pass through the $\hat{n}$ axis, then
$\vec{r}^{\;\text{opt}}$ lies along this axis$.$

{\bf Case 2.3: }If the chord defined by $\hat{a}_{1}^{0}$ and $\hat{a}%
_{2}^{0}$ and the chord defined by $\hat{a}_{1}^{1}$ and
$\hat{a}_{2}^{1}$ are parallel to, equidistant from, and on
either side of the equatorial plane perpendicular to $\hat{n},$
then $\vec{r}^{\;\text{opt}}$ lies in that plane.

If the conditions of cases 2.2 and 2.3 both hold, then
$\vec{r}^{\;\text{opt}}$ lies at the centre of the Bloch ball. This
corresponds to Alice submitting the completely mixed state. An
example of such a protocol is provided in Fig.~9. Although in the
example of this figure the two chords point in orthogonal
directions, this is not necessary, it is only necessary that they
not be parallel.

The proofs of the results of this section are presented in
Appendix B.

\subsubsection{One set contains one element and one set contains two
elements $(n_{0}=1,n_{1}=2)$}
\label{VID2}

We now assume that one of the sets $\left\{ \psi _{k}^{0}\right\}
_{k=1}^{n_{0}}$ and $\left\{ \psi _{k}^{1}\right\}
_{k=1}^{n_{1}}$ has only a single element while the other has
two. Without loss of generality we may assume that the single
element set is the $b=0$ set, and we denote its unique element by
$\left| \psi ^{0}\right\rangle .$ So in order to unveil a bit
value of $1$ Alice can announce either $k=1$ or $k=2$ and must
then pass Bob's test for $\left| \psi _{k}^{1}\right\rangle ,$
while to unveil a bit value of $0$ Alice has no choice but to
pass a test for the state $\left|
\psi ^{0}\right\rangle .$

We first consider the case where $\left\langle \psi ^{0}|\psi
_{1}^{1}\right\rangle =\left\langle \psi ^{0}|\psi
_{2}^{1}\right\rangle .$
This corresponds to case 1 of section~6.4.1 in the limit that $\hat{a}%
_{1}^{0}$ and $\hat{a}_{2}^{0}$ converge to a single point
$\hat{a}^{0}$ representing $\left| \psi ^{0}\right\rangle .$
There is a family of optimal solutions of the form
\[
\vec{r}^{\;\text{opt}}=\frac{\hat{a}^{0}+\hat{a}_{1}^{1}}{\left| \hat{a}^{0}+%
\hat{a}_{1}^{1}\right| }+\lambda \hat{d}_{1},\mbox{ for }0\le \lambda \le 2.
\]
This family corresponds to the points on the chord of the Bloch
ball that is parallel to the chord defined by
$\hat{a}_{1}^{1},\hat{a}_{2}^{1}$ and that passes through the
point on the surface of the ball that is equidistant between
$\hat{a}^{0}$ and $\hat{a}_{1}^{1}$. The BC protocol illustrated
in Fig.~10 is an example of this case. The case $\left\langle \psi
^{0}|\psi _{1}^{1}\right\rangle \ne \left\langle \psi ^{0}|\psi
_{2}^{1}\right\rangle $ corresponds to case 2 of section~6.4.1
in the limit that $\hat{a}_{1}^{0}$ and $\hat{a}_{2}^{0}$
converge to the point $\hat{a}^{0}$. In this limit, we find that
$\left| \vec{r}^{\;\text{max}}\right| >1$. Consequently,
\[
\vec{r}^{\;\text{opt}}=\frac{\hat{a}^{0}+\hat{a}_{1}^{1}}{\left| \hat{a}^{0}+%
\hat{a}_{1}^{1}\right| }.
\]

\subsubsection{Both sets contain one element $(n_{0}=1,n_{1}=1)$}
\label{VID3}

We now assume there is only a single element in both of the sets,
and denote each of these states by $\left| \psi ^{b}\right\rangle
.$ Thus to unveil a bit value of $b$ Alice must pass Bob's test
for $\left| \psi ^{b}\right\rangle .$ Consider first the
possibility that $\left| \psi ^{0}\right\rangle $ and $\left|
\psi ^{1}\right\rangle $ are orthogonal. In this case, no matter
what $\rho $ Alice submits, her probability of unveiling either
bit is strictly 1/2.

When $\left| \psi ^{0}\right\rangle $ and $\left| \psi
^{1}\right\rangle $ are not orthogonal, the situation corresponds
to case 2 of section~6.4.1, in the limit that $\hat{a}_{1}^{b}$
and $\hat{a}_{2}^{b}$ converge to a
single point $\hat{a}^{b}$ for both values of $b$. In this limit we again find $\left| \vec{r}^{%
\;\text{max}}\right| >1.$ Consequently,
\[
\vec{r}^{\;\text{opt}}=\frac{\hat{a}^{0}+\hat{a}^{1}}{\left| \hat{a}^{0}+%
\hat{a}^{1}\right| }.
\]
An example is presented in Fig.~8.

\begin{figure}[htbp]
\centerline{\epsfig{file=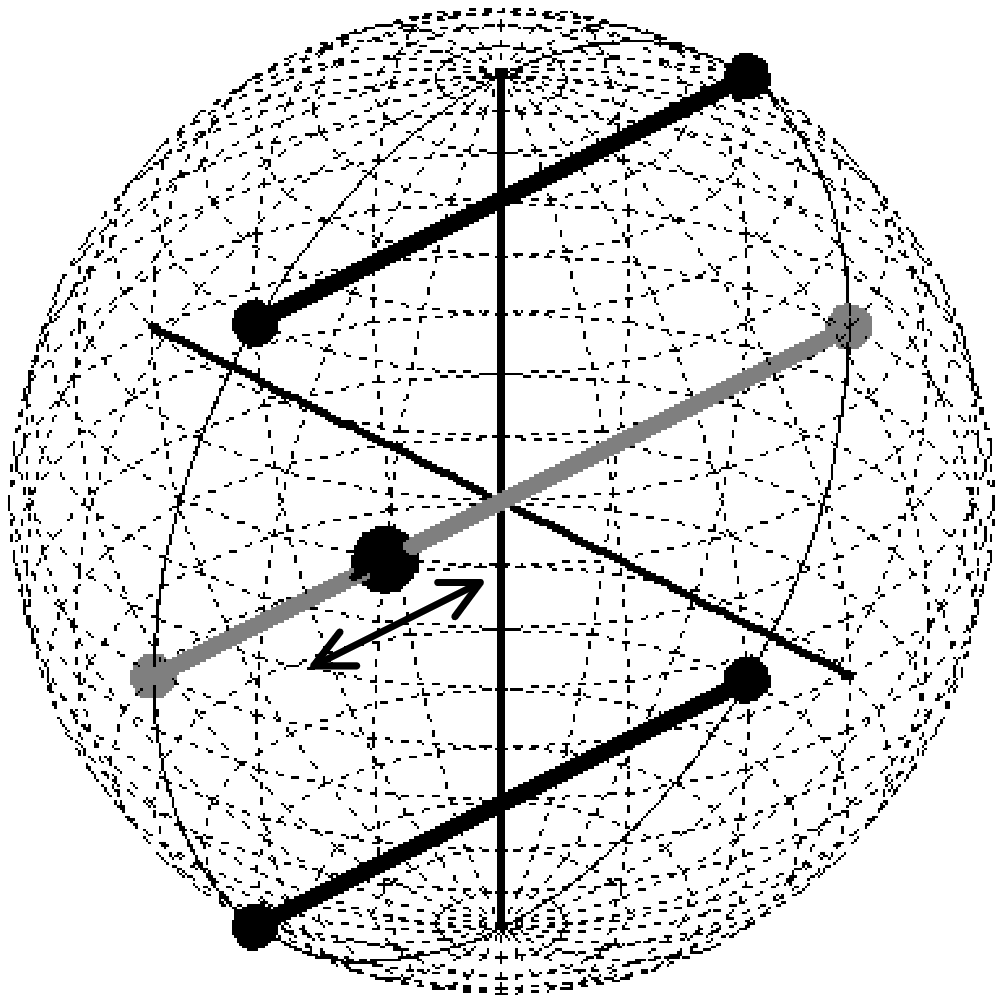,width=50mm,bbllx=5cm,bblly=9cm,bburx=18cm,bbury=21.5cm,clip=}}
\fcaption{ An illustration of a BC\ protocol of the form
proposed by Aharonov {\em et al.}~\cite{Aharanov}. The two sets of
states are given by Eq.~(\ref {Aharonovstates}) with $\theta =\pi
/8.$ There is a family of optimal density operators lying along
the chord indicated in grey.}
\end{figure}

\section{Applications of the results}
\label{VII}
\noindent
 These results can be applied to the generalized BB84 BC protocol proposed by
Aharonov {\em et al.}~\cite{Aharanov}. The protocol is defined by
the following states, from which an honest Alice chooses uniformly
\begin{eqnarray}
\left| \psi _{1}^{0}\right\rangle  &=&\left| \theta \right\rangle ,\text{ }%
\left| \psi _{2}^{0}\right\rangle =\left| -\theta \right\rangle ,  \nonumber
\\
\left| \psi _{1}^{1}\right\rangle  &=&\left| \pi /2-\theta \right\rangle
,\left| \psi _{2}^{1}\right\rangle =\left| \pi /2+\theta
\right\rangle ,
\label{Aharonovstates}
\end{eqnarray}
where $\left| \theta \right\rangle =\cos \theta \left|
0\right\rangle +\sin
\theta \left| 1\right\rangle $ and $\theta $ is some fixed angle satisfying $%
0<\theta \le \frac{\pi }{4}.$ The sets of states associated with
bits $0$ and $1$ describe parallel chords on the Bloch ball, as
depicted in Fig.~7. We therefore have an instance of case 1 of
section~6.4.1. It follows that an
optimal strategy for Alice is to simply submit $\left| +\right\rangle =\frac{%
1}{\sqrt{2}}\left( \left| 0\right\rangle +\left| 1\right\rangle
\right) $ and tell Bob to test for $\left| \psi
_{1}^{b}\right\rangle $, where $b$ is
the bit she wishes to unveil. Another is to submit $\left| -\right\rangle =%
\frac{1}{\sqrt{2}}\left( \left| 0\right\rangle -\left| 1\right\rangle
\right) $ and to tell Bob to test for $\left| \psi _{2}^{b}\right\rangle .$
So Alice does not need to make use of entanglement in this case.
The most general optimal strategy is for Alice to submit $\rho
=w\left| +\right\rangle \left\langle +\right| +(1-w)\left|
-\right\rangle
\left\langle -\right| ,$ realize the convex decomposition $\left\{ \left(
w,\left| +\right\rangle \left\langle +\right| \right) ,\left(
(1-w),\left| -\right\rangle \left\langle -\right| \right)
\right\} ,$ and tell Bob to test for $\left| \psi
_{1}^{b}\right\rangle \left( \left| \psi _{2}^{b}\right\rangle
\right) $ upon obtaining the outcome $\left| +\right\rangle
\left( \left| -\right\rangle \right) .$ Alice's maximum
probability of unveiling whatever bit she desires is
\[
P_{U}^{\text{max}}=\frac{1}{2}\left( 1+\sin 2\theta \right) .
\]

Previously, the best known upper bound on this probability was
\[P_{U}\le \frac{1}{2}\left( 1+\frac{1}{\cos ^{2}2\theta }\left(
\sqrt{1+2\cos ^{2}2\theta }-1\right) \right) ,\] as can be
inferred from the results in
section 5 of Ref.~\cite{Aharanov}. In the case of $\theta =\pi /8,$ we find $%
P_{U}^{\text{max}}=\frac{1}{2}+\frac{1}{2\sqrt{2}}\simeq
.\,85355,$ while the previous best bound was $P_{U}\le
\frac{\sqrt{8}-1}{2}\simeq \allowbreak .91421.$

We can now compare this with Bob's maximal probability of
estimating Alice's commitment correctly. If Alice follows the
honest protocol for committing a bit $b,$ she chooses uniformly
between $\left| \psi _{1}^{b}\right\rangle $ and $\left| \psi
_{2}^{b}\right\rangle $ and submits a system in this state.
Bob must therefore discriminate between density operators $\rho _{0}$ and $%
\rho _{1}$ defined by $\rho _{b}=\frac{1}{2}\sum_{k=1}^{2}\left| \psi
_{k}^{b}\right\rangle \left\langle \psi _{k}^{b}\right| .$ His
maximum probability of doing so is given by Eq.~(\ref{Helstrom}),
in this case,
\[
P_{E}^{\text{max}}=\frac{1}{2}\left( 1+\cos 2\theta \right) .
\]
It is worth noting that the quantity $\frac{1}{2}${\rm Tr}$\left|
\rho _{0}-\rho _{1}\right| $ appearing in Eq.~(\ref{Helstrom}) is
simply half the Euclidean distance between the points in the
Bloch ball representing $\rho _{0}$ and $\rho _{1}.$

\begin{figure}[htbp]
\centerline{\epsfig{file=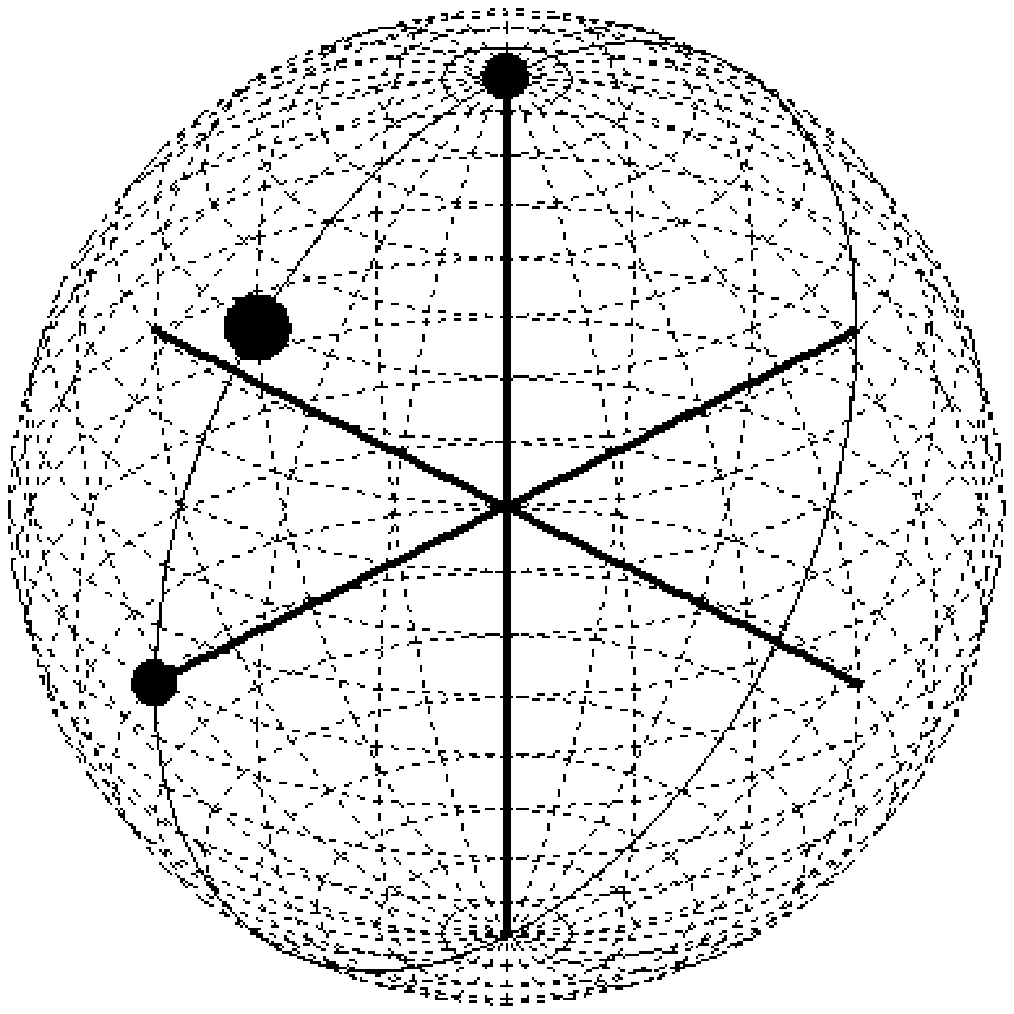,width=50mm,bbllx=5cm,bblly=9cm,bburx=18cm,bbury=21.5cm,clip=}}
\fcaption{ An illustration of a BC\ protocol of the form
specified in Eq.~(\ref {2 pure states}), with $\gamma =\pi /4.$
The optimal density operator is indicated by the large black
sphere. BC\ protocols of this form achieve the same trade-off
between concealment and bindingness as those of the form proposed
by Aharonov {\em et al.}}
\end{figure}

>From the above expression for $P_{U}^{\max }$ and $P_{E}^{\max
},$ we can conclude that there is a trade-off between these
quantities of the form
\begin{equation}
(P_{U}^{\text{max}}-1/2)^{2}+(P_{E}^{\text{max}}-1/2)^{2}=1/4.
\label{Anaronovtradeoff}
\end{equation}
At $\theta =0,$ $P_{E}^{\text{max}}=1$ and
$P_{U}^{\text{max}}=1/2,$ so that there is no concealment against
Bob, but perfect bindingness against Alice
(since $P_{U}^{\text{max}}=1/2$ for an honest Alice). At $\theta =\pi /4,$ $%
P_{E}^{\text{max}}=1/2$ and $P_{U}^{\text{max}}=1,$ so that the
roles of Alice and Bob are reversed. The only choice of $\theta $
leading to a `fair'
protocol is $\theta =\pi /8.$ In this case, $P_{E}^{\text{max}}=P_{U}^{\text{%
max}}=\frac{1}{2}+\frac{1}{2\sqrt{2}}.$

Our results also imply that the same trade-off between
$P_{U}^{\text{max}}$ and $P_{E}^{\text{max}}$ can be achieved
with the most simple imaginable BC protocol, namely one wherein
Alice submits to Bob one of two non-orthogonal states.
Specifically, to commit a bit $b,$ an honest Alice sends Bob a
qubit in the state $\left| \psi ^{b}\right\rangle ,$ where
\begin{eqnarray}
\left| \psi ^{0}\right\rangle  &=&\left| 0\right\rangle ,  \nonumber \\
\text{ }\left| \psi ^{1}\right\rangle  &=&\left| \gamma \right\rangle ,
\label{2 pure states}
\end{eqnarray}
where $\gamma $ is some fixed angle satisfying $0<\gamma \le \pi
/2.$ An
example of this protocol is illustrated in Fig.~8. This is an instance where $%
n_{0}=1$ and $n_{1}=1$, which was considered in section~6.4.3.
One can infer from the results of that section that Alice's
optimal strategy is to submit the state $\left| \gamma
/2\right\rangle $ and to announce whatever bit she wishes to
unveil. It is straightforward to verify that this protocol has
the same properties as the one described above.

It is easy to understand the equivalence of these protocols geometrically. $%
P_{U}^{\max }$ is proportional to the cosine of the angular
separation of the endpoints of the polytopes (chords or points)
representing the sets of states an honest Alice chooses from.
Meanwhile, $P_{E}^{\max }$ is proportional to the Euclidean
distance between the midpoints of these polytopes. It is easy to
see from Figs. 7 and 8 that if the endpoints have the same
angular separation, then the midpoints have the same Euclidean
separation.

Interestingly, it turns out that any protocol satisfying the
conditions of cases 2.2 and 2.3 of section~6.4.1 {\em also}
yields exactly the same trade-off between $P_{U}^{\text{max}}$
and $P_{E}^{\text{max}}$. Specifically, one can use any protocol
of the form
\begin{eqnarray}
\left| \psi _{1}^{0}\right\rangle  &=&\left| \theta ,0\right\rangle ,\text{ }%
\left| \psi _{2}^{0}\right\rangle =\left| -\theta ,0\right\rangle ,
\nonumber \\
\left| \psi _{1}^{1}\right\rangle  &=&\left| \pi /2-\theta ,\phi
\right\rangle ,\left| \psi _{2}^{1}\right\rangle =\left| \pi /2+\theta
,-\phi \right\rangle ,  \label{non-orthog Aharonov}
\end{eqnarray}
where $\left| \theta ,\phi \right\rangle =\cos \theta \left|
0\right\rangle +e^{i\phi }\sin \theta \left| 1\right\rangle $ and
$\theta $ and $\phi $ are fixed angles satisfying $0<\theta \le
\frac{\pi }{4},$ $0<\phi \le \pi /2.$ Fig.~9 depicts an example of
such a protocol. Geometrically, $P_{U}^{\max }$ is no longer
given by the angle between the endpoints of the two polytopes
representing the states an honest Alice chooses from but rather
the angle between the endpoints of these polytopes and the
closest endpoints of the polytopes representing the elements of
the convex decomposition that Alice realizes. This ensures that $P_{U}^{\max }$ is the same as for the protocols discussed above. The only difference is
that Alice's optimal strategy in this case requires the use of
entanglement.

\begin{figure}[thbp]
\centerline{\epsfig{file=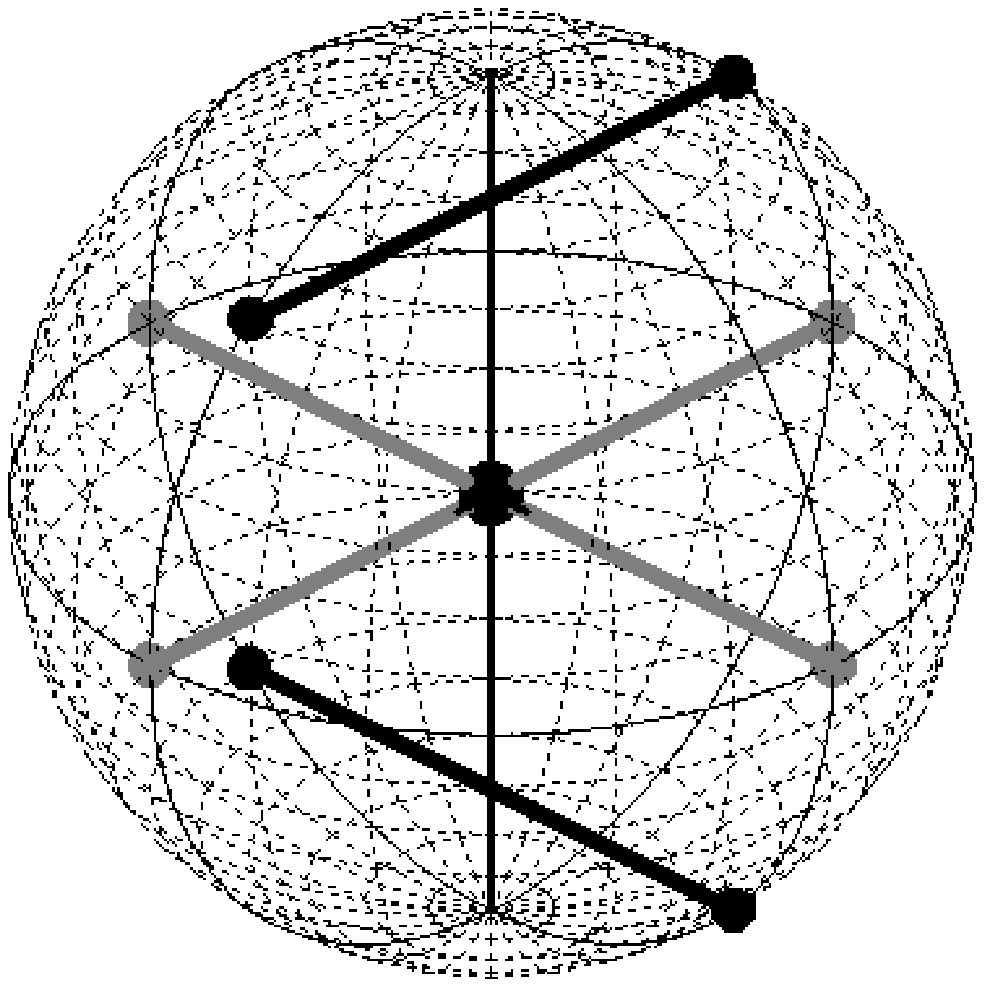,width=50mm,bbllx=5cm,bblly=9cm,bburx=18cm,bbury=21.5cm,clip=}}
\fcaption{ An illustration of a BC protocol of the form
specified in Eq.~(\ref {non-orthog Aharonov}), with $\theta =\pi
/8$ and $\phi =\pi .$ The optimal density operator in this case
lies at the origin. Depending on which bit Alice desires to
unveil, she realizes the convex decomposition parallel to one or
the other of the two chords.}
\end{figure}

Finally, we consider a protocol wherein there is a single state
associated with committing a bit $0$ but two states associated
with committing bit $1,$ specifically,
\begin{eqnarray}
\left| \psi ^{0}\right\rangle  &=&\left| 0\right\rangle ,  \nonumber \\
\left| \psi _{1}^{1}\right\rangle  &=&\left| \alpha \right\rangle ,\left|
\psi _{2}^{1}\right\rangle =\left| -\alpha \right\rangle ,
\label{optimalqubitstates}
\end{eqnarray}
where $\alpha $ is some fixed angle satisfying $0<\alpha \le
\frac{\pi }{2}.$ An example of this protocol is provided in
Fig.~10. It is of the form described in section~6.4.2, with
$\left\langle \psi ^{0}|\psi _{1}^{1}\right\rangle =\left\langle
\psi ^{0}|\psi _{2}^{1}\right\rangle .$ From the results of that
section, we can infer that there are a family of optimal coherent
attacks of the following form. Alice submits a density operator
of the form $\rho =w\left|
\alpha /2\right\rangle \left\langle
\alpha /2\right| +(1-w)\left| -\alpha /2\right\rangle \left\langle -\alpha
/2\right| .$ If she decides to try to unveil bit $1,$
she realizes the convex decomposition $\{ \left( w,\left|\alpha
/2\right\rangle\left\langle \alpha /2\right|
\right),$ $\left( (1-w),\left| -\alpha /2\right\rangle \left\langle
-\alpha /2\right| \right)\} ,$ and upon obtaining the outcome
$\left|
\alpha /2\right\rangle
\left( \left| -\alpha /2\right\rangle \right) $ tells Bob to test
for $\left| \alpha \right\rangle
\left( \left| -\alpha \right\rangle \right) .$ Alice's maximum probability
of unveiling whatever bit she desires in this case is

\[
P_{U}^{\max }=\frac{1}{2}+\frac{1}{2}\cos \alpha .
\]
Meanwhile, from Eq.~(\ref{Helstrom}), we can infer that Bob's
maximum probability of correctly estimating Alice's commitment is
\[
P_{E}^{\max }=\frac{1}{2}+\frac{1}{2}\sin ^{2}\alpha .
\]
The trade-off between $P_{U}^{\max }$ and $P_{E}^{\max }$ is
\begin{equation}
2\left( P_{U}^{\max }-\frac{1}{2}\right) ^{2}+\left( P_{E}^{\max }-\frac{1}{2%
}\right) =\frac{1}{2}.  \label{optimalqubittradeoff}
\end{equation}
A `fair' protocol has $P_{U}^{\max }=P_{E}^{\max }=\frac{1}{2}+\frac{\sqrt{5}%
-1}{4}\simeq \allowbreak .\,80902.$

 From a comparison of the trade-offs (\ref
{Aharonovstates}) and (\ref{optimalqubittradeoff}), it is easy to
verify that a protocol which uses the states
(\ref{optimalqubitstates}) achieves, for a given bindingness (a
given $P_{U}^{\max })$, a concealment that is greater (and thus a
$P_{E}^{\max }$ that is smaller) than the concealment
that can be achieved in a protocol using the states (\ref{Aharonovstates}), (%
\ref{2 pure states}) or (\ref{non-orthog Aharonov}).

This point is easy to see geometrically. We compare this last protocol with the protocol defined by the states
(\ref{2 pure states}) for simplicity. From the examples provided in Figs. 8 and 10 it is easy
to visualize the fact that if the endpoints of the polytopes defined by the two
protocols have the same angular separation, the midpoints do {\em
not }have the same Euclidean separation -- the separation
is smaller for a BC protocol defined by the states (\ref{optimalqubitstates})%
$.$

An obvious question to ask at this point is whether the trade-off
relation of Eq.~(\ref{optimalqubittradeoff}) is optimal, in the
sense that the concealment against Bob is maximized for a given
bindingness against Alice. Elsewhere~\cite{SpekkensRudolph2} we
show that it {\em is }optimal among a certain class of protocols
(which includes the generalized BB84 protocols) that can be
implemented using a single qubit. We also show that a better
trade-off can be achieved with a BC protocol that makes use of a qu{\em trit}, that is, a three-level
system. The protocol we suggest in Ref.~\cite{SpekkensRudolph2} is
not a generalized BB84 BC protocol; however, an equivalent
protocol that {\em is} of the generalized BB84 form has been
proposed by Ambainis~\cite{Ambainis}.

\begin{figure}[htbp]
\centerline{\epsfig{file=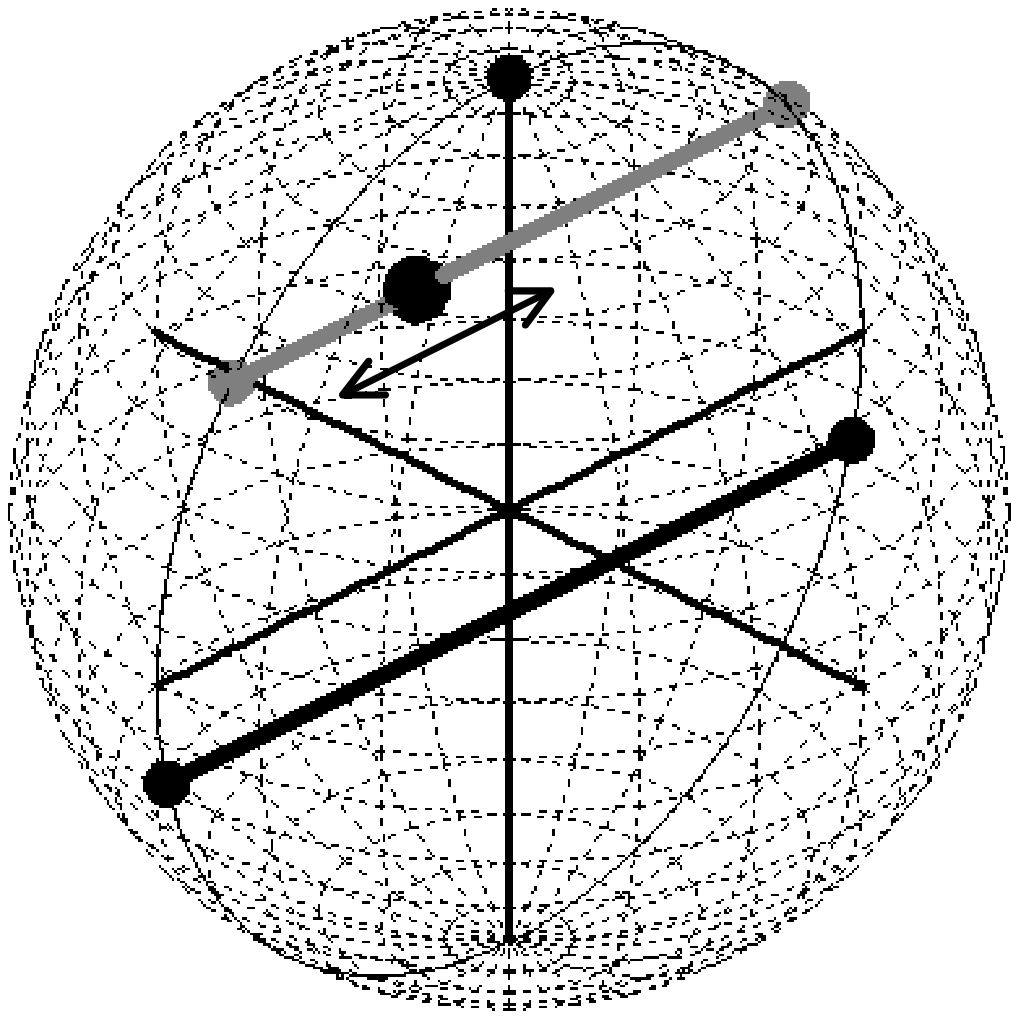,width=50mm,bbllx=5cm,bblly=9cm,bburx=18cm,bbury=21.5cm,clip=}}
\fcaption{ An illustration of a BC\ protocol of the form
specified in Eq.~(\ref {optimalqubitstates}), with $\alpha
=\arccos \left( (\sqrt{5}-1)/2\right) $. There is a family of
optimal density operators lying along the chord indicated in
grey. BC\ protocols of this form achieve a better trade-off
between concealment and bindingness than those of the form
proposed by Aharonov {\em et al.}}
\end{figure}

\section{Conclusions}
\label{VIII}
\noindent
We have formulated the problem of optimizing coherent attacks on
Generalized BB84 BC protocols in terms of a theorem of
Hughston, Jozsa and Wootters. We have found that there is a
mapping between this problem and one of state estimation.
Specifically, we have shown that the convex decomposition that is optimal for successfully preparing
one of a set of states is related in a simple way to the POVM
measurement that is optimal for discriminating among certain
transformations of these states.

We have identified Alice's optimal coherent attack for a class of
generalized BB84 BC protocols that can be implemented using a
single qubit. From these results we have determined the degree of
bindingness that can be achieved in the BC protocol proposed by
Aharonov {\it et al.}, improving upon the best previous upper
bound. This enables us to identify the trade-off between the
degree of concealment and the degree of bindingness for this
protocol. It has also led us to identify several qubit protocols
that achieve the same trade-off as the proposal of Aharonov {\it
et al. }, as well as a qubit protocol that achieves a better
trade-off.

In optimizing over Alice's strategies, we have relied on the
Bloch ball representation of quantum states. This provides a
convenient geometrical picture of a coherent attack. Although
this representation can be generalized to higher
dimensions~\cite{generalizedBlochball}, it is unlikely that the geometrical pictures acheivable in this way will be as intuitive. In
any event, there remain many questions to be answered even for
qubit protocols, for which this approach is likely to provide
some insight. For instance, one can use it to consider qubit BC
protocols that are {\em not }generalizations of the BB84 BC
protocol.

In another paper~\cite{SpekkensRudolph2}, we determine the optimal
coherent attack in a class of BC protocols that is larger than
the set of generalized BB84 protocols. However, the problem of
determining the optimal trade-off between concealment and
bindingness from among {\em all }BC protocols remains open.

Beyond their relevance to bit commitment, coherent attacks are
interesting as an example of what might be considered a
fundamental task in quantum information processing, namely, the
preparation of quantum states at a remote location.  One can
define many variants of this task, depending on whether the
parties at the two locations are cooperative or adversarial, and
depending on the available resources, such as the number of
classical or quantum bits that can be exchanged, and the amount
of prior entanglement the parties share.  Bennett {\it et
al.}~\cite{Bennettetal} have recently considered remote state
preparation in the case of cooperative parties who share prior
entanglement and a classical channel.  In the type of remote
state preparation we have considered in this paper, the parties
are adversarial and although Alice makes use of a quantum
channel, she does so at a time prior to knowing which state she
is supposed to prepare.

It seems to us that the primitive of remote state preparation,
construed in its most general sense, may be as fundamental as the
primitive of state estimation and just as significant for the
purposes of determining what sorts of information processing
tasks can be successfully implemented with quantum systems.
The mapping discussed above between state estimation and the
particular type of remote state preparation considered in this
paper suggests that there may be other connections between these
two problems.  In future work, we hope to explore this analogy in
more detail.

\nonumsection{Acknowledgments}
\noindent
We gratefully acknowledge J. E. Sipe for useful discussions. This
work was supported by the Natural Sciences and Engineering
Research Council of Canada, the Austrian Science Foundation FWF,
and the TMR programs of the European Union Project No.
ERBFMRXCT960087.

\nonumsection{References}

\appendix
\noindent

We here provide the proof of optimality of the convex
decomposition specified by Eq.~(\ref{optdecomp4fixedrho}). First,
we establish the applicability of Jaynes' rule, defined in
Eq.~(\ref{Jaynesrule}), to the probabilities in the optimal convex
decomposition. This requires showing that the optimal
decomposition is an extremal decomposition with the number of
positively-weighted elements equal to the rank of $\rho .$

This is trivial to see for a pure $\rho .$ We now demonstrate it
for an impure $\rho .$ Because we have assumed that the $\left\{
\psi _{k}\right\} _{k=1}^{n}$ are linearly independent, they span
the whole 2D Hilbert space, and because $\rho $ has rank 2, its
support is the 2D Hilbert space. Thus, the $\left\{ \psi
_{k}\right\} _{k=1}^{n}$ are linearly independent and have a span
that is equal to the support of $\rho ,$ which, as shown in
section~5.1 by the mapping to the state estimation problem, is
sufficient to establish that the optimal convex decomposition is
an extremal decomposition. It was shown in section~4 that the
number of positively-weighted elements in the optimal convex
decomposition is less than or equal to $n.$ In the present case,
$n=2,$ so this number must be less than or equal to 2. However,
since $\rho $ is impure, every convex decompositions of $\rho $
has {\em at least }2 elements receiving non-zero
probability. Thus, the number must be precisely 2, which is the rank of $%
\rho .$

Jaynes' rule provides a formula for the probabilities in a convex
decomposition of $\rho $ in terms of $\rho $ and the elements in
the decomposition. In terms of Bloch vectors, it has the form
\[
q_{k}=\frac{1}{2}\frac{1-\left| \vec{r}\right| ^{2}}{1-\vec{r}\cdot \hat{s}%
_{k}},
\]
where we have written $\hat{s}_{k}$ rather than $\vec{s}_{k}$
since the elements of the optimal decomposition, being pure, can
be represented by unit Bloch vectors. Substituting this
expression, together with the constraint that
$\vec{r}=q_{1}\hat{s}_{1}+q_{2}\hat{s}_{2}$ into Eq.~(\ref
{P_Ub Bloch}), we can write $P_{Ub}$ entirely in terms of $\hat{s}_{1},$%
\begin{eqnarray*}
P_{Ub} &=&\frac{1}{2}\left( 1+q_{1}\left( \hat{a}_{1}\cdot \hat{s}%
_{1}\right) +\left( \hat{a}_{2}\cdot \vec{r}\right) -q_{1}\left( \hat{a}%
_{2}\cdot \hat{s}_{1}\right) \right)  \\
&=&\frac{1}{2}\left( 1+\hat{a}_{2}\cdot \vec{r}\right)
+\frac{1}{4}\left(
1-\left| \vec{r}\right| ^{2}\right) \frac{\left( \left( \hat{a}_{1}-\hat{a}%
_{2}\right) \cdot \hat{s}_{1}\right) }{\left( 1-\vec{r}\cdot \hat{s}%
_{1}\right) }.
\end{eqnarray*}
Rather than varying this quantity with respect to $\hat{s}_{1},$
we vary
with respect to an unnormalized vector $\vec{s}_{1},$ taking $\hat{s}_{1}=%
\vec{s}_{1}/\left| \vec{s}_{1}\right| ,$ where $\left| \vec{s}_{1}\right| =%
\sqrt{\vec{s}_{1}\cdot \vec{s}_{1}}.$ Setting $\delta P_{E}(\hat{s}_{1})=0$
and making use of the fact that $\delta \left| \vec{s}_{1}\right|
=\delta
\sqrt{\vec{s}_{1}\cdot \vec{s}_{1}}=\delta \vec{s}_{1}\cdot \hat{s}_{1},$ we
find that the optimal $\hat{s}_{1}$ satisfies
\[
\left( 1-\hat{s}_{1}\cdot \vec{r}\right) \left( \hat{a}_{1}-\hat{a}%
_{2}\right) +\left( \hat{s}_{1}\cdot \left(
\hat{a}_{1}-\hat{a}_{2}\right)
\right) \left( \vec{r}-\hat{s}_{1}\right) =0.
\]
By assumption, $\left| \vec{r}\right| \ne 1$ (since $\rho $ is
impure). It follows that $\left( 1-\hat{s}_{1}\cdot
\vec{r}\right) \ne 0,$ and $\left(
\vec{r}-\hat{s}_{1}\right) \ne \vec{0}.$ Since it is also the case that $%
\left( \hat{a}_{1}-\hat{a}_{2}\right) \ne \vec{0},$ we infer that $\hat{s}%
_{1}\cdot \left( \hat{a}_{1}-\hat{a}_{2}\right) \ne 0$. Taking
the dot product of this equation with $\hat{a}_{1}+\hat{a}_{2},$
we find
\[
\left( \vec{r}-\hat{s}_{1}\right) \cdot \left( \hat{a}_{1}+\hat{a}%
_{2}\right) =0.
\]
Consequently, the solutions that extremize $P_{Ub}$ are of the
form
\[
\hat{s}_{1\pm }^{\text{ext}}=\vec{r}+L_{\pm }\left( \vec{r}\right) \hat{d},
\]
where $\hat{d}$ is given in Eq.~(\ref{d}). The constraint
$\left|
\hat{s}_{1\pm }^{\text{ext}}\right| =1$ implies that $L_{\pm }\left( \vec{r}%
\right) $ have the form specified in Eq.~(\ref{Lpm}). This result implies that
$\hat{a}_{1}\cdot \hat{s}_{1}^{\text{ext}}=\hat{a}_{2}\cdot
\hat{s}_{2}^{\text{ext}}$, which allows one to simplify the
expression for $P_{Ub}$ provided in Eq.~(\ref{P_Ub Bloch}).
Plugging $\hat{s}_{1\pm }^{\text{ext}}$ into the resulting
expression yields
\[
P_{Ub}=\frac{1}{2}\left( 1+\hat{a}_{1}\cdot \vec{r}+L_{\pm }\left( \vec{r}%
\right) \left( \hat{a}_{1}\cdot \hat{d}\right) \right) .
\]
Since the coefficient of $L_{\pm }\left( \vec{r}\right) $ is positive, and $%
L_{+}\left( \vec{r}\right) \ge L_{-}\left( \vec{r}\right) ,$ the maximum $%
P_{Ub}$ occurs for $\hat{s}_{1+}^{\text{ext}},$ while the minimum
occurs for $\hat{s}_{1-}^{\text{ext}}$. Thus, the optimal
$\hat{s}_{1}$ given $\vec{r}$ is
\[
\hat{s}_{1}^{\text{opt}}=\vec{r}+L_{+}\left( \vec{r}\right) \hat{d}.
\]
The constraint $\vec{r}=\sum_{k}q_{k}\vec{s}_{k}$ then implies
that
\[
\hat{s}_{2}^{\text{opt}}=\vec{r}+L_{-}\left( \vec{r}\right) \hat{d}.
\]
This establishes what we set out to prove.

\appendix
\noindent

We here present the proofs of the results of section~6.4.1.

{\bf Proof for case 1.} The parallel condition is equivalent to $\hat{d}_{0}=%
\hat{d}_{1}$ which implies that $L_{0+}=L_{1+},$ so that Eq.~(\ref{P_U Bloch}%
) becomes
\[
P_{U}=\frac{1}{2}+\frac{1}{4}\left( \vec{r}+L_{0+}\left( \vec{r}\right) \hat{%
d}_{0}\right) \cdot \left( \hat{a}_{1}^{0}+\hat{a}_{1}^{1}\right)
.
\]
This is maximized for $\vec{r}+L_{0+}\left( \vec{r}\right) \hat{d}_{0}=\frac{%
\hat{a}_{1}^{0}+\hat{a}_{1}^{1}}{\left| \hat{a}_{1}^{0}+\hat{a}%
_{1}^{1}\right| },$ which implies that $\vec{r}^{\;\text{opt}}$ can
be any vector of the form specified in Eq.~(\ref{ropt
case1}).~$\Box $

{\bf Proof for case 2. }Starting from Eq.~(\ref{P_U Bloch}) , we extremize $%
P_{U}$ with respect to variations in $\vec{r}$ by setting $\delta P_{U}(\vec{%
r})=0.$ Using the fact that $\delta r=\delta \sqrt{\vec{r}\cdot \vec{r}}%
=\delta \vec{r}\cdot \vec{r}/r,$ and
\[
\delta L_{b+}\left( \vec{r}\right) =-\left( \frac{\vec{r}+L_{b+}\left( \vec{r%
}\right) \hat{d}_{b}}{\vec{r}\cdot \hat{d}_{b}+L_{b+}\left( \vec{r}\right) }%
\right) \cdot \delta \vec{r},
\]
we find that the extremal $\vec{r}$ satisfy
\begin{equation}
\sum_{b=0}^{1}\left( \hat{a}_{1}^{b}-\left( \hat{d}_{b}\cdot \hat{a}%
_{1}^{b}\right) \left( \frac{\vec{r}+L_{b+}\left( \vec{r}\right) \hat{d}_{b}%
}{\vec{r}\cdot \hat{d}_{b}+L_{b+}\left( \vec{r}\right) }\right)
\right) =0.
\label{optimal r}
\end{equation}
  We now introduce the notation
\begin{equation}
x_{0} =\vec{r}\cdot \hat{d}_{1}^{\perp }, \mbox{ }
x_{1} =\vec{r}\cdot \hat{d}_{0}^{\perp }, \mbox{ }
x_{2} =\vec{r}\cdot \hat{n},
\end{equation}
where $\hat{d}_{1}^{\perp },\hat{d}_{0}^{\perp }$ and $\hat{n}$
are defined
in Eq.~(\ref{nonorth Bloch basis}). Making use of the fact that $\vec{r}%
=\left( \vec{r}\cdot \hat{d}_{0}\right) \hat{d}_{0}+\left( \vec{r}\cdot \hat{%
d}_{0}^{\perp }\right) \hat{d}_{0}^{\perp }+\left( \vec{r}\cdot \hat{n}%
\right) \hat{n},$ we have $\left( \frac{\vec{r}+L_{0+}\left( \vec{r}\right)
\hat{d}_{0}}{\vec{r}\cdot \hat{d}_{0}+L_{0+}\left( \vec{r}\right) }\right) =%
\hat{d}_{0}+\frac{x_{1}\hat{d}_{0}^{\perp }+x_{2}\hat{n}}{\sqrt{%
1-x_{1}^{2}-x_{2}^{2}}},$ which together with $\hat{a}_{1}^{0}=\left( \hat{a}%
_{1}^{0}\cdot \hat{d}_{0}\right) \hat{d}_{0}+\left(
\hat{a}_{1}^{0}\cdot
\hat{d}_{0}^{\perp }\right) \hat{d}_{0}^{\perp }+\left( \hat{a}_{1}^{0}\cdot
\hat{n}\right) \hat{n}$ yields
\begin{eqnarray*}
&&\hat{a}_{1}^{0}-\left( \hat{d}_{0}\cdot \hat{a}_{1}^{0}\right)
\left(
\frac{\vec{r}+L_{0+}\left( \vec{r}\right) \hat{d}_{0}}{\vec{r}\cdot \hat{d}%
_{0}+L_{0+}\left( \vec{r}\right) }\right)= \\
&&\left( \hat{a}_{1}^{0}\cdot \hat{d}_{0}^{\perp }\right) \hat{d}%
_{0}^{\perp }+\left( \hat{a}_{1}^{0}\cdot \hat{n}\right) \hat{n}-\frac{%
\left( \hat{d}_{0}\cdot \hat{a}_{1}^{0}\right) }{\sqrt{1-x_{1}^{2}-x_{2}^{2}}%
}\left( x_{1}\hat{d}_{0}^{\perp }+x_{2}\hat{n}\right) .
\end{eqnarray*}
An analogous result holds for $b=1.$ Plugging these expressions
into Eq.~(\ref {optimal r}) and taking the dot product with each
of $\hat{d}_{0},\hat{d}_{1} $ and $\hat{n},$ we obtain the set of
equations
\begin{eqnarray*}
0&=&\left( \hat{a}_{1}^{1}\cdot \hat{d}_{1}^{\perp }\right) \sqrt{%
1-x_{0}^{2}-x_{2}^{2}}-\left( \hat{a}_{1}^{1}\cdot
\hat{d}_{1}\right) x_{0},
\\
0&=&\left(\hat{a}_{1}^{0}\cdot \hat{d}_{0}^{\perp }\right) \sqrt{%
1-x_{1}^{2}-x_{2}^{2}}-\left( \hat{a}_{1}^{0}\cdot
\hat{d}_{0}\right) x_{1},
\\
0&=&\left( \left( \hat{a}_{1}^{0}+\hat{a}_{1}^{1}\right) \cdot
\hat{n}\right)
\sqrt{1-x_{0}^{2}-x_{2}^{2}}\sqrt{1-x_{1}^{2}-x_{2}^{2}} \\
&-&\left( \hat{a}_{1}^{0}\cdot \hat{d}_{0}\right) \sqrt{1-x_{0}^{2}-x_{2}^{2}}%
x_{2}-\left( \hat{a}_{1}^{1}\cdot \hat{d}_{1}\right)
\sqrt{1-x_{1}^{2}-x_{2}^{2}}x_{2}.
\end{eqnarray*}
The values of $x_{0},x_{1}$ and $x_{2}$ that maximize $P_{U},$ denoted by $%
x_{0}^{\text{max}},x_{1}^{\text{max}}$ and $x_{2}^{\text{max}},$
are easily
seen to be those given by Eq.~(\ref{x opt}). These define $\vec{r}^{\;\text{max}%
}$ through Eq.~(\ref{r max}).

If $\left| \vec{r}^{\;\text{max}}\right| \le 1$, then it
corresponds to the optimal density operator. If $\left|
\vec{r}^{\;\text{max}}\right| >1,$ then there is no extremum of
$P_{U}$ inside the Bloch ball and the optimal density operator
must be represented by a point on the boundary of the ball. Such
a point corresponds to a pure state. Consequently there is no
freedom in the convex decomposition Alice realizes, and all that
she must decide is what state to tell Bob to test for. If she
tells him $\left| \psi
_{k}^{0}\right\rangle $ when she wishes to unveil a bit value of $0$ and $%
\left| \psi _{k^{\prime }}^{1}\right\rangle $ when she wishes to unveil a
bit value of $1,$ then in terms of Bloch vectors, her probability
of unveiling the bit of her choosing is

\begin{eqnarray*}
P_{U} &=&\frac{1}{4}\left( 1+\hat{r}\cdot \hat{a}_{k}^{0}\right) +\frac{1}{4}%
\left( 1+\hat{r}\cdot \hat{a}_{k^{\prime }}^{1}\right) \\
&=&\frac{1}{2}+\frac{1}{4}\hat{r}\cdot \left( \hat{a}_{k}^{0}+\hat{a}%
_{k^{\prime }}^{1}\right) ,
\end{eqnarray*}
where we write $\hat{r}$ to emphasize that we are varying over
pure density operators. The vector
$\hat{r}=\frac{\hat{a}_{k}^{0}+\hat{a}_{k^{\prime }}^{1}}{\left|
\hat{a}_{k}^{0}+\hat{a}_{k^{\prime }}^{1}\right| }$ clearly
maximizes $P_{U}.$ In our notational convention, $\hat{a}_{1}^{0}$ and $\hat{%
a}_{1}^{1}$ are the closest pair of Bloch vectors from the two
sets, so Alice should choose $k=k^{\prime }=1.$ It follows that
the optimal density operator is represented by the Bloch vector
defined in Eq.~(\ref{r opt}).

Note that it may occur that $\hat{a}_{2}^{0}$ and
$\hat{a}_{2}^{1}$ are as close to one another as
$\hat{a}_{1}^{0}$ and $\hat{a}_{1}^{1},$ that is, it may occur
that there is no unique `closest' pair of Bloch vectors. However,
in this case one will not find $\left|
\vec{r}^{\;\text{max}}\right| >1.$ The
reason is as follows. If one {\em did} find $\left| \vec{r}^{\;\text{max}%
}\right| >1,$ then the optimal $\vec{r}$ would have to be a pure
state.
However, since the pure states associated with the Bloch vectors $\frac{\hat{%
a}_{1}^{0}+\hat{a}_{1}^{1}}{\left|
\hat{a}_{1}^{0}+\hat{a}_{1}^{1}\right| }$
and $\frac{\hat{a}_{2}^{0}+\hat{a}_{2}^{1}}{\left| \hat{a}_{2}^{0}+\hat{a}%
_{2}^{1}\right| }$ would yield the same $P_{U},$ any mixture of
these would also yield this $P_{U}.$ This in turn would imply
that there existed a solution with $\left|
\vec{r}^{\;\text{max}}\right| <1.~\Box $

{\bf Proof for case 2.1}. Since $\hat{a}_{1}^{0}$, $\hat{a}_{2}^{0}, \hat{a}%
_{1}^{1}$ and $\hat{a}_{2}^{1}$ all lie in a plane,
$\hat{a}_{k}^{b}\cdot
\hat{n}$ is independent of $b$ and $k.$ In this case, we find $x_{0}^{\text{%
max}}=\hat{a}_{1}^{1}\cdot \hat{d}_{1}^{\perp },$ $x_{1}^{\text{max}}=\hat{a}%
_{1}^{0}\cdot \hat{d}_{0}^{\perp }$ and $x_{2}^{\text{max}}=\hat{a}%
_{1}^{0}\cdot \hat{n}.${\em \ }That this corresponds to the point
of intersection can be verified from the parametric equations for
the lines containing the two chords.$~\Box $

{\bf Proof for case 2.2}. If the chord defined by $\hat{a}_{1}^{b}$ and $%
\hat{a}_{2}^{b}$ passes through the $\hat{n}$ axis, then it must lie in the
plane of $\hat{d}_{b}$ and $\hat{n},$ so that $\hat{a}_{k}^{b}\cdot \hat{d}%
_{b}^{\perp }=0.$ It follows that
$x_{0}^{\text{max}}=x_{1}^{\text{max}}=0,$ and thus
$\vec{r}^{\;\text{max}}=x_{2}^{\text{max}}\hat{n}.$ Since $\left|
x_{2}^{\text{max}}\right| \le 1,$ we know that $\left| \vec{r}^{\;\text{max}%
}\right| \le 1,$ so that $\vec{r}^{\;\text{opt}}=\vec{r}^{\;\text{max}}=x_{2}^{%
\text{max}}\hat{n}.~\Box $

{\bf Proof for case 2.3}. The case being considered corresponds to $\hat{n}%
\cdot \left( \hat{a}_{k}^{0}+\hat{a}_{k}^{1}\right) =0.$ We must consider
the two possibilities $\left| \vec{r}^{\;\text{max}}\right| \le 1$
and $\left|
\vec{r}^{\;\text{max}}\right| >1.$ In the former, $\vec{r}^{\;\text{opt}}=\vec{r}%
^{\;\text{max}},$ while in the latter $\vec{r}^{\;\text{opt}}=\frac{\hat{a}%
_{1}^{0}+\hat{a}_{1}^{1}}{\left| \left( \hat{a}_{1}^{0}+\hat{a}%
_{1}^{1}\right) \right| }.$ Either way, the condition
$\hat{n}\cdot \left(
\hat{a}_{k}^{0}+\hat{a}_{k}^{1}\right) =0$ implies that $\vec{r}^{\;\text{opt}%
}\cdot \hat{n}=0.~\Box $


\begin{thebibliography}{000}
\bibitem{Localitycaveat}  A discussion of the progression towards the full impossibility theorem can be found in \cite{Bub}. Note that this result applies to bit commitment protocols that are based solely on quantum mechanics and for which the laboratories of the two parties have spatial dimensions that are small compared to their spatial separation. If these assumptions are relaxed, then relativistic constraints allow one to securely implement a form of bit commitment, as demonstrated by A. Kent \cite{Kent}.
\end{thebibliography}
\end{document}